\documentclass[a4paper,11pt]{article}
\pdfoutput=1 

\usepackage{jheppub} 

\usepackage[T1]{fontenc} 
\usepackage{amsmath,graphicx,amsfonts, mathrsfs,amssymb}
\usepackage{amsmath,epsfig}
\usepackage{amssymb,amsfonts}
\usepackage{latexsym}
\usepackage{subfigure}
\usepackage{pdfsync}
\usepackage{braket}
\usepackage{bm}
\def\be{\begin{equation}}
\def\ee{\end{equation}}
\def\bea{\begin{eqnarray}}
\def\eea{\end{eqnarray}}
\def \t0{{\tau_0}}
\title{Entanglement Properties of Boundary State and Thermalization}

\author[]{Wu-zhong Guo}

\affiliation[]{Physics Division, National Center for Theoretical Sciences,\\National Tsing-Hua University, Hsinchu 30013, Taiwan}

\emailAdd{wzguo@cts.nthu.edu.tw}

\abstract{We discuss the regularized boundary state $e^{-\tau_0 H}\ket{B}_a$ on two aspects in both 2D CFT and higher dimensional free field theory. One is its entanglement and correlation properties, which exhibit exponential decay in 2D CFT, the parameter $1/\tau_0$ works as a mass scale. The other concerns with its time evolution, i.e.,  $e^{-i tH}e^{-\tau_0 H}\ket{B}_a$. We investigate the Kubo-Martin-Schwinger (KMS) condition on correlation function of \emph{local} operators to detect the thermal properties. Interestingly we find the correlation functions in the  initial state  $e^{-\tau_0 H}\ket{B}_a$ also partially satisfy the KMS condition. In the limit $t\to \infty$, the correlators will exactly satisfy the KMS condition.  We generally analyse  quantum quench by a pure state and obtain some constraints on the possible form of 2-point correlation function in the initial state if assuming they satisfies KMS condition in the final state . As a byproduct we find in an large $\tau_0$ limit  the thermal property of 2-point function in $e^{-\tau_0 H}\ket{B}_a$ also appears. }

\begin{document}
\maketitle
\flushbottom
\section{Introduction}
Boundary state appears in conformal field theory (CFT) defined in finite spacetime\cite{Cardy1}\cite{Cardy2}. It is one of the special elements in Hilbert space of CFT. Without regularization the norm of boundary state is not well defined, its energy is also expected to be divergent, but the real space entanglement is vanishing\cite{Miyaji:2014mca}. \\
To study such a special state suitable regularization are necessary. For a boundary state $\ket{B}_a$, usually we regularize it by introducing a scale cut-off $\tau_0$ and define a new state $\ket{B}^{\tau_0}_{a}\equiv e^{-\tau_0 H}\ket{B}_a$. This state can be taken as the initial state to study the global quantum quench in 2D conformal field theory\cite{Calabrese:2006rx}\cite{Calabrese:2007rg}\cite{Cardy:2015xaa}. In this process the correlation functions of local operators will approach to thermal ones after long time, the regularized cut-off $\tau_0$ gains a real physical meaning which is found to be associated with the temperature. Please refer to the review \cite{Calabrese:2016xau} and references in it for more recent progresses.\\
In this paper we would like to study two aspects on the regularized boundary state. \\
Firstly, we focus on the entanglement properties of this state, which is directly related to the behavior of correlation functions. Usually the correlation functions in this state exponentially decay for spatial separation, the parameter $\tau_0$ controls the decay rate, working as a mass scale. We show a cluster property for spatially separated two bounded operators. With this we use the Bell inequality to study the quantum entanglement in the boundary state. But we also found an example in higher dimension the correlator in regularized boundary state may not exponentially decay. \\
Secondly, we also consider the time evolution of $\ket{B}^{\tau_0}_{a}$ and discuss the thermal properties after long time evolution. We will  investigate  the Kubo-Martin-Schwinger (KMS) condition on correlation functions of local operators\cite{Kubo}\cite{Martin}\cite{Hagg}. For two operators the KMS condition means
\be\label{KMSintroduction}
\omega(A \tau_{i\beta}(B))=\omega(B A),\quad \text{with}\quad  \tau_z (A)\equiv e^{i z H}A e^{-i z H},
\ee
and $F(z)\equiv \omega(A\tau_{z}(B))$ is analytic in the region $0<\text{Im}(z)<\beta $ (if $\beta >0$).Our motivation to use KMS condition is that it concerns with the correlation properties of operators rather than the state itself, this may give us more insights on the possible  relation between the initial correlation behavior and final thermal property. Of course  the state after time evolution can't be a thermal state, since our initial state is a pure state. More precisely the KMS condition we will use is not for global operators but restricted to local operators, which means the distance between operators should be not too large.  This is similar as a local version of KMS condition \cite{Buchholz:2001qj}, in which one could compare the state with a global KMS-state by means of local operators. We will comment on this more in section \ref{generalanalysis}.\\
Indeed we find the initial state $\ket{B}^{\tau_0}_a$ do hide some information on the thermal properties. An evidence is that the 2-point function in $\ket{B}^{\tau_0}_a$ also satisfies the KMS relation (\ref{KMSintroduction}), but $F(z)$ is non-analytic. In the free field theory the hidden information becomes almost obvious when considering the correlator of $a_{\bm{k}}$ and $a^{\dagger}_{\bm{k}}$. We will analyse the role of the time evolution and see why the thermal properties appear.\\
We also try to answer the following question. Start with a state $\Ket{\Psi}$ in 2D CFT, which is not an eigenstate of Hamiltonian, if assuming the correlation functions in the final state satisfy KMS condition, what is the possible constraints on the correlation function in initial state $\Ket{\Psi}$.  We mainly discuss the 2-point connected correlation function,
\begin{eqnarray}
&&C_{t}(x_1,x_2)\equiv \bra{\Psi(t)}O(w_1,\bar w_1)O(w_2,\bar w_2)\ket{\Psi(t)}\nonumber \\
&&- \bra{\Psi(t)}O(w_1,\bar w_1)\ket{\Psi(t)}\bra{\Psi(t)}O(w_2,\bar w_2)\ket{\Psi(t)},\nonumber
\end{eqnarray}
where $\ket{\Psi(t)}=e^{-it H}\ket{\Psi}$. The general form of $C_{t}(x_1,x_2)$ is
\be
C_{t}(x_1,x_2)=M(t)T(w_1-w_2,\bar w_1-\bar w_2)+N(t),
\ee
where $M(t\to\infty)=\text{Constant}\ne 0$ and $N(t\to\infty)=0$,  $T(w_1-w_2,\bar w_1-\bar w_2)$ satisfies KMS condition. \\
With this condition we find $M(t=0)$ would always exponentially decay or a constant for large spatial separation, while $N(t=0)$ could be  exponentially decaying or polynomially decaying depending on the details of the states. Therefore, the final state could be like a  thermal state even if the initial state has long distance correlation.

As a byproduct  we find when assuming two operators are very close or equally the parameter $\tau_0$ is large, the initial boundary state $\ket{B}^{\tau_0}_a$ also exhibits thermal properties. This may be associated with the subsystem thermalization which has many processes recently. \\

We will review the boundary state in CFT in section \ref{review}. Then we study the correlators in regularized boundary state $\ket{B}^{\tau_0}_a$ in section \ref{section2}. We derive an estimation on cluster of two bounded operators, based on which we obtain an upper bound on violation of Bell inequality. Section \ref{timeevolution} is devoted to discuss the KMS condition for the final state. In section \ref{general} we generally analyse the reason why the thermal properties appear in the final state, and discuss some constraints on the possible form of 2-point correlation function in the initial state if assuming the final state satisfies the KMS condition. The last section is the conclusion.

\section{Review on boundary state}\label{review}
In this section we review definitions and some basic properties of boundary states in CFTs. We mainly focus on the 2D rational CFTs and free massless scalar field in (d+1)-dimensional flat spacetime.
\subsection{Boundary state in 2D CFT}
Following the discuss in \cite{Cardy1}, we consider CFT defined on a finite cylinder of circumference $R$ and length  $L$. The boundary conditions $a,b$ are imposed on the edges of the cylinder. The coordinates of the cylinder are $w=x+i\tau$ and $\bar w=x-i\tau$, the time $\tau$ is along the cylinder, with $0\le \tau\le L$. In this case the boundary conditions $a,b$ can be described by  boundary states. By a coordinate transformtion
\be\label{cylindertoz}
z=e^{-2\pi i w/R},
\ee
the cylinder is mapped onto the $z$ plane, in which two boundaries $a,b$ becomes two concentric circles, representing boundary states $\Ket{a},\Ket{b}$ in radial quantization.\\
As shown in \cite{Cardy2}, to keep the boundary invariant one must impose a condition for stress energy tensor on the boundary, for the finite cylinder,we have
\be
T^{cyl}(x,0)=\bar T^{cyl}(x,0)\quad \text{and}\quad T^{cyl}(x,L)=\bar T^{cyl}(x,L).
\ee
On the $z$ plane these conditions become
\be
T^{pl}(\xi)\xi^2=\bar T^{pl}(\bar \xi)\bar \xi^2,
\ee
We obtain constraints on the boundary states $\Ket{a},\Ket{b}$,
\be \label{boundarystateEOM}
(L_n-\bar L_{-n})\Ket{a}(\Ket{b})=0,
\ee
where $L_n,\bar L_{n}$ are the  Virasoro generators on the $z$ plane. \\
In any Verma modules $\mathcal{V}_j\bigotimes \mathcal{\bar V}_{\bar j}$, the equations (\ref{boundarystateEOM}) have solutions,
which are some special states called Ishibashi states  $\ket{j}\rangle$ \cite{Ishibashi},
\be\label{Ishibashistate}
\ket{j}\rangle\equiv \sum_{N}\ket{j,N}\otimes \ket{\bar j,N},
\ee
with $j=\bar j$, $\ket{j,N}$ are the descendant states at  level-$N $ in the Verma module $\mathcal{V}_j\bigotimes \mathcal{\bar V}_{\bar j}$.\\
A physical boundary state (or Cardy state), denoted by $\Ket{B}_a$  is a linear combination of Ishibashi states. $\Ket{B}_a$ should satisfy the consistent conditions of partition function of the finite cylinder, or the so-called Cardy equation \cite{Cardy1}. Generally, we have
\be
\ket{B}_a=\sum_j C_a^j \ket{j}\rangle,
\ee
in rational CFT the sum is finite. For the diagonal minimal models the coefficients $C_a^j$ are derived in \cite{Cardy1}, the Cardy states are
\be\label{cardystate}
\ket{B}_a=\sum_j \frac{S^i_a}{(S^i_0)^{1/2}}\ket{i}\rangle,
\ee
where $S^i_a$ is the modular matrix element of the Virasoro characters under modular transformation $\mathcal{S}$.\\
The operator $L_0+\bar L_0$  generates the dilations $(z,\bar z) \to \lambda(z,\bar z)$, which is proportional to Hamiltonian in radial quantization. $L_0-\bar L_0$ is the generator of rotation in the $z$ plane. It is obvious that the Ishibashi states (\ref{Ishibashistate}) are invariant under rotation, but not under dilation transformation. It means  the Ishibashi states are space-translation invariant in the finite cylinder, but will change under time evolution.

\subsection{Free field theory boundary state}
The boundary state can be generalized to higher dimensional CFT. The free massless scalar field in (d+1)-dimensional spacetime is the simplest example. We could impose Neumann ($+$) or Dirichlet($-$) boundary conditions to keep the conformal symmetry. The corresponding boundary states can be expressed in the Fock space as
\be \label{freescalarboundarystate}
\ket{B}_{\pm}=e^{\pm \frac{1}{2}\int d^dk a_{\bm{k}}^\dagger a_{-\bm{k}}^\dagger}\ket{0},
\ee
where $a_{\bm{k}}^\dagger$ is the creation operator. Notice that the boundary states are space-translation invariant. One could check this by using the space-translation generator $\bm{P}$,
\be
\bm{P}=\int d^dp\  \bm{p}\ a_{\bm{p}}^\dagger a_{\bm{p}},
\ee
and directly calculate $e^{-i a \bm{P}}\ket{B}_{\pm}$. But under time evolution, which is generated by the Hamiltonian $H$,
\be
H=\int d^d k\ |\bm{k}|a_{\bm{k}}^\dagger a_{\bm{k}},
\ee
the states $\ket{B}_{\pm}$ will change. For simplicity we only consider $d=3$ below.\\

\subsection{Regularize the boundary state}\label{regboundarysection}
The norm of the Ishibashi state defined by (\ref{Ishibashistate}) is divergent, since the representation on $\mathcal{V}_j\bigotimes \mathcal{\bar V}_{\bar j}$ is infinite dimension. One could also directly check the product of the free scalar boundary state (\ref{freescalarboundarystate}) is  not-convergent. In this paper we would like to study a regularized boundary state
\be\label{regularizedstate}
\ket{B}^{\tau_0}_{a}=e^{-\tau_0 H}\ket{B}_{a},
\ee
where $H$ is the Hamiltonian of CFT, $\tau_0$ is a  positive  constant. Several comments are in order.  At first the state (\ref{regularizedstate}) is still space-translation invariant, since $[\bm{P},H]=0$. Secondly, in the path-integral formalism the correlation functions for $\ket{B}^{\tau_0}_{a}$ could be evaluated as path-integral on strip shape of width $2\tau_0$ in Euclidean spacetime  with operators inserted. Thirdly,  $\tau_0$ is not just a regularization parameter, it has physical meaning if we consider the correlation function or time evolution of such state as we will show below. \\
Let's see the regularized free scalar boundary state, which could be expressed as
\be
\ket{B}^{\tau_0}_{\pm}=\mathcal{N}e^{\pm \frac{1}{2}\int d^d k e^{-2\tau_0 E_k} a^\dagger_{\bm k}a^\dagger_{-\bm k}}\ket{0},
\ee
where $E_k=|\bm{k}|$, $\mathcal{N}$ is the normalization constant. We have the following properties,
\begin{eqnarray}\label{twocacorrelation}
&&~_{\pm}^\t0\!\bra{B}a^\dagger_{\bm k}a_{\bm p}\ket{B}^\t0_{\pm}=\frac{1}{e^{4\tau_0 E_k}-1}\delta(\bm k- \bm p)\label{cacorrelation} \\
&&~_{\pm}^\t0\!\bra{B}a_{\bm k}a_{\bm p}\ket{B}^\t0_{\pm}=\pm\frac{e^{2\tau_0 E_k}}{e^{4\tau_0 E_k}-1}\delta(\bm k+\bm p) \\
&&~_{\pm}^\t0\!\bra{B}a_{\bm k}a^\dagger_{\bm p}\ket{B}^\t0_{\pm}=\frac{e^{4\tau_0 E_k}}{e^{4\tau_0 E_k}-1}\delta(\bm k- \bm p) \\
&&~_{\pm}^\t0\!\bra{B}a^\dagger_{\bm k}a^\dagger_{\bm p}\ket{B}^\t0_{\pm}=\pm\frac{e^{2\tau_0 E_k}}{e^{4\tau_0 E_k}-1}\delta(\bm k+\bm p).
\end{eqnarray}
The correlation function for $a^\dagger_{\bm k}a_{\bm p}$ (\ref{cacorrelation}) is same as the one in thermal field theory with  $\beta\equiv 1/T=4 \tau_0$. More importantly, it is time-independent,  but the correlation function of $a_{\bm k}a_{\bm p}$ will change under time evolution. This implies the state (\ref{regularizedstate}) has some information on thermal field theory with  $\beta=4\tau_0$. Notice that the correlation function of $a^\dagger_{\bm k}a_{\bm p}$ (\ref{cacorrelation}) is independent on the boundary state we choose, but the $a_{\bm k}a_{\bm p}$ correlation function is related to the boundary condition.

\section{Entanglement properties of boundary state}\label{section2}
Boundary state is one of the special states in CFT on its entanglement properties. It is argued in paper \cite{Miyaji:2014mca} the real space entanglement of the Cardy state should be vanishing. But as we can see in the definition of Ishibashi state (\ref{Ishibashistate})the left and right-moving sectors are maximally entangled. It is still not clear whether these two phenomenons have some relation. In this section we will discuss the correlation function  and real space entanglement of the regularized boundary state (\ref{regularizedstate}). Our tool is the Bell inequality for two spacelike regions.
\subsection{Correlation functions in regularized boundary state}\label{sectioncorrelation}
In 2D CFT we are interested in the correlation function
\begin{eqnarray}
&& ~_a^\t0\!\bra{B}O(w_1,\bar w_1)O(w_2,\bar w_2)...O(w_n,\bar w_n)\ket{B}^{\tau_0}_{a}\nonumber \\
&&=~_a\!\bra{B}e^{-\tau_0 H}O(w_1,\bar w_1)O(w_2,\bar w_2)...O(w_n,\bar w_n)e^{-\tau_0 H}\ket{B}^{\tau_0}_{a},
\end{eqnarray}
which is the correlation function in an infinite long strip of width $2\tau_0$ in the Euclidean spacetime with the boundary conditions on the edge ($\tau_E=-\tau_0,\tau_0$) corresponding to the boundary state $\ket{B}_a$\cite{Calabrese:2006rx}. The coordinate of the strip is denoted by $w=x+i \tau$, by a conformal map
\be\label{conformalmap}
w(z)=\frac{2\tau_0}{\pi}\log (z)-i \tau_0 \quad \text{and }\quad \bar w(\bar z)=\frac{2\tau_0}{\pi}\log (\bar z)+i \tau_0,
\ee
the strip is mapped to the upper half-plane(UHP) (Im $z>0$). One- and two-point correlation function has been obtained in \cite{Calabrese:2006rx}. For one-point correlation function,
\be
~_a^\t0\!\bra{B}O(w,\bar w)\ket{B}^{\tau_0}_{a}=w'(z)^{-h} \bar w' (\bar z)^{-h} \langle O(z,\bar z)\rangle_{\text{UHP}},
\ee
in which
\be
\langle O(z,\bar z)\rangle_{\text{UHP}}=A^O_a [\text{Im} (z-\bar z)]^{-2h},
\ee
where $A^O_a$ is a universal constant depending both on the field $O$ and boundary condition $a$ \cite{CardyLewellen}. We have
\be\label{onepoint}
~_a^\t0\!\bra{B}O(w,\bar w)\ket{B}^{\tau_0}_{a}=A^O_a \Big(\frac{\pi}{4\tau_0} \frac{1}{\cosh[(w-\bar w)\pi/4\tau_0]}\Big)^{2h}.
\ee
Two-point function on UHP has the following general form \cite{Cardy2},
\be
\langle O(z_1,\bar z_1)O(z_2,\bar z_2)\rangle_{\text{UHP}}= (z_{12} z_{\bar 1\bar 2})^{-2h}x^{-2h}F(x),
\ee
where $x=z_{1\bar 1 }z_{2\bar 2}/z_{1\bar 2}z_{2\bar 1}$ is the cross ration of $z_1,z_2$ and their images $\bar z_1,\bar z_2$, $F(x)$ depends only on $x$ and can be expanded by conformal blocks. We would like to consider the case $x\sim 0$, which means the horizonal distance (denoted by $\rho$) between the two points approaches to infinity, i.e., $\rho\to \infty$. In this limit we assume the 2-point function in an ``extraordinary transition'', which means the leading terms are
\be\label{UHPfirst}
\langle O(z_1,\bar z_1)O(z_2,\bar z_2)\rangle_{\text{UHP}}\sim\langle O(z_1,\bar z_1)\rangle_{\text{UHP}} \langle O(z_2,\bar z_2)\rangle_{\text{UHP}}(1+M \frac{1}{\rho^{\eta_{||}}}),
\ee
where the number $\eta_{||}$ is called by the surface exponent, $M$ is associated with the distances of the two points from real axis.

We could obtain
\be\label{crossration}
x=\frac{e^{\pi(w_1+w_2)/2\tau_0}+e^{\pi(w_1+\bar w_2)/2\tau_0}+e^{\pi(\bar w_1+ w_2)/2\tau_0}+e^{\pi(\bar w_1+\bar w_2)/2\tau_0}}{e^{\pi(w_1+w_2)/2\tau_0}+e^{\pi(w_1+\bar w_1)/2\tau_0}+e^{\pi(w_2+\bar w_2)/2\tau_0}+e^{\pi(\bar w_1+\bar w_2)/2\tau_0}}.
\ee
In general $F(x)$ depends on the details of CFT and boundary conditions. But when $x\sim0$ and $x\sim 1$, $F(x)$ is expected to have universal forms, since the identity channel will mainly contribute to $F(x)$. For $x\sim 0$, $z_{1\bar 1},z_{2\bar 2}\sim 0$ , which means the correlation between the points and their images will be the leading contribution, $F(x)\simeq (A^O_a)^2+...$\footnote{In general, the leading contribution is $(A^O_a)^2x^{h_b}$, where $h_b$ is the boundary scaling dimension of the boundary operator to which $O$ couples. Here we assume it is the identity $h_b=0$.}. Oppositely for $x\sim 1$, the two points will be far away from the boundary, as a result $F(x)\simeq 1$.\\
The two-point function on the strip would be
\be\label{2D2point}
~_a^\t0\!\bra{B}O(w_1,\bar w_1)O(w_2,\bar w_2)\ket{B}^{\tau_0}_{a}=\Big(\frac{\pi}{4\tau_0}\Big)^{4h}\Big[x\sinh(\frac{\pi(w_1-w_2)}{4\tau_0})\sinh(\frac{\pi(\bar w_1-\bar w_2)}{4\tau_0})\Big]^{-2h}F(x).
\ee
 For $O(w_1,\bar w_1)$ and $O(w_2,\bar w_2)$ are spacelike, take a special case $w_1=\bar w_1=0$ and $w_2=\bar w_2=L$, and  assume $L\gg \tau_0$. We have  $x=1/(\cosh[L\pi/4\tau_0])^2 \sim 0$, in this limit  by using (\ref{UHPfirst}) with $\rho= e^{L\pi/(2\tau_0)}-1$ we obtain
\be
~_a^\t0\!\bra{B}O(0,0)O(L,0)\ket{B}^{\tau_0}_{a}\simeq~_a^\t0\!\bra{B}O(0,0)\ket{B}^{\tau_0}_{a}~_a^\t0\!\bra{B}O(L,0)\ket{B}^{\tau_0}_{a}(1+M e^{-\frac{\pi L \eta_{||}}{2\tau_0}}),
\ee
Consider the connected two-point function,
\be
C(x_1,x_2)\equiv ~_a^\t0\!\bra{B}O(0,0)O(L,0)\ket{B}^{\tau_0}_{a}-~_a^\t0\!\bra{B}O(0,0)\ket{B}^{\tau_0}_{a}~_a^\t0\!\bra{B}O(L,0)\ket{B}^{\tau_0}_{a},
\ee
 we have
\be\label{connect2D}
C(x_1,x_2)\simeq M ~_a^\t0\!\bra{B}O(0,0)\ket{B}^{\tau_0}_{a}~_a^\t0\!\bra{B}O(L,0)\ket{B}^{\tau_0}_{a} e^{-\frac{\pi L \eta_{||}}{2\tau_0}}.
\ee
This means that one could always find some constant $M'$ such that the connected two-point function $C(x_1,x_2)\le M' e^{-\frac{\pi\eta_{||} d(x_1,x_2)}{2\tau_0}}$, where $d(x_1,x_2)$ is the distance between two points. The  spatial 2-point functions  are exponential decay in the regularized boundary state (\ref{regularizedstate}). \\

For the scalar field theory we have a little different result. We would like to consider the two-point correlation function of scalar field $\phi(x,t=0)$ in the regularized boundary state, i.e.,$~_{\pm}^\t0\!\bra{B}\phi(x,0)\phi(y,0)\ket{B}^{\tau_0}_{\pm}$. By using (\ref{twocacorrelation}), we obtain
\be
~_{\pm}^\t0\!\bra{B}\phi(x,0)\phi(y,0)\ket{B}^{\tau_0}_{\pm}=\frac{1}{(2\pi)^3}\int d^3 k \frac{1}{2E_k} \frac{e^{2\tau_0 E_k}\pm 1}{e^{2\tau_0 E_k}\mp 1}e^{i \bm{k}\cdot (\bm{x-y})}.
\ee
The one-point correlation function is vanishing, since $~_{\pm}^\t0\!\bra{B} a_{\bm{k}}(a^\dagger_{\bm{k}})\ket{B}^{\tau_0}_{\pm}$. The connected correlation function is
\begin{eqnarray}
C_{\pm}(x,y)&=&\frac{1}{(2\pi)^3}\int d^3 k \frac{1}{2E_k} \frac{e^{2\tau_0 E_k}\pm 1}{e^{2\tau_0 E_k}\mp 1}e^{i \bm{k}\cdot (\bm{x-y})}\nonumber \\
&\propto &\frac{1}{r} \int_{-\infty}^{+\infty}dk \frac{e^{2\tau_0 k}\pm 1}{e^{2\tau_0 k}\mp 1}e^{-i k r},
\end{eqnarray}
where $r=|x-y|$. For $r\gg \tau_0$ we have
\begin{eqnarray}\label{poly}
&&C_{+}(x,y)\propto \frac{1}{\tau_0 r}\coth(\frac{\pi r}{2\tau_0})\ \sim\  \tau_0^{-1}r^{-1} \\
&&C_{-}(x,y)\propto \frac{1}{\tau_0r\sinh(\frac{\pi r}{2\tau_0})}\ \sim\  \tau_0^{-1}r^{-1}e^{-\frac{\pi r}{2\tau_0}}.
\end{eqnarray}
This result shows in higher dimension ($d>2$) the spatial correlation function in boundary state may not be exponential decay\footnote{For 2D scalar field  $\phi$ is not a primary operator.In the vacuum its correlation function  $\langle\phi(x)\phi(y)\rangle\propto\log (x-y)^2$, which is not polynomial decay. But its derivative $\partial_z \phi(z)$ is primary operator with conformal dimension $h=1$, and has the correlation function $\langle \partial_z \phi(z) \partial_w\phi(w)\rangle \propto \frac{1}{(z-w)^2}$. We could check the correlation function of this operator in regularized Neumann boundary state,
\begin{eqnarray}
~_{+}^{\tau_0}\!\bra{B}\partial_x\phi(x,0)\partial_y\phi(y,0)\ket{B}^{\tau_0}_{+}=\frac{1}{2\pi}\int_{-\infty}^{+\infty}dk k \frac{e^{2\tau_0 k}+ 1}{e^{2\tau_0 k}- 1}e^{i k r}\sim \tau_0^{-2} \text{csch }(\frac{\pi r}{2\tau_0})\sim  \tau_0^{-2} e^{-\frac{\pi r}{2\tau_0}},
\end{eqnarray}
if $r\gg \tau_0$. }. The different behaviors of the correlation would lead to distinct physical phenomenons. However, for Neumann boundary state not all the operators are polynomially decaying, such as the operator $\pi(x)$,
\begin{eqnarray}
~_{+}^\t0\!\bra{B}\pi(x,0)\pi(y,0)\ket{B}^{\tau_0}_{+}&=&\frac{1}{(2\pi)^3}\int d^3 k \frac{E_k}{2} \frac{e^{2\tau_0 E_k}- 1}{e^{2\tau_0 E_k}+ 1}e^{i \bm{k}\cdot (\bm{x-y})}\nonumber \\
&&\propto \frac{1}{r}\int_{-\infty}^{+\infty}dk k^2 \frac{e^{2\tau_0 k}- 1}{e^{2\tau_0 k}+ 1}e^{ik r}\sim  e^{-\frac{\pi r}{2 \tau_0}}.
\end{eqnarray}
This is consistent with the definition of Neumann boundary state, $\pi(x)\ket{B}_{+}=0$. So there exists operator that is polynomially decaying in higher dimension, different from the 2D CFT.

\subsection{Energy gap and cluster property}

There is a secret relation between a non-vanishing mass gap and exponential decay of correlation function in vacuum state both in non-relativistic and relativistic quantum theory. There exists models with unique vacuum and exponential decay of correlation function but without a mass gap. But the inverse statement for quantum lattice models is proved to be true under some conditions\cite{Fredenhagen}\cite{Exponential1}\cite{Exponential2}. Their discussions mainly focus on the clustering properties of vacuum state. Here our discussion is different, for CFT the correlation in the vacuum  should be power-law decay, but in the regularized boundary state the exponential decay appears. \\

The energy density of (\ref{regularizedstate}) $\langle T_{tt}\rangle_B=\frac{\pi c}{24 (2\tau_0)^2}$ is a time-independent constant, which can be considered as the Casimir energy density in the strip of width $2\tau_0$\cite{HWJ}. So the total energy of the regularized boundary state is also a constant. As we have show in section \ref{sectioncorrelation} the correlations in the state (\ref{regularizedstate}) can be derived by respective correlations in a strip.  In \cite{Cardy3}\cite{Cardy86} Cardy shows the energy gap of the excited states for a strip with varied boundary conditions by using two-point functions. \\
Let's recall the two-point function
\be\label{twopointversion1}
\langle O(0,0)O(L,0)\rangle_{\text{strip}}=~_a^\t0\!\bra{B}O(L,0)O(0,0)\ket{B}^{\tau_0}_{a}\sim 1+ M e^{-\frac{\pi L \eta_{||}}{2\tau_0}},
\ee
for $L\gg \tau_0$. We have two alternative ways to see the two-point function in a strip. First, one could take the $\tau$-direction as the Euclidean time. In this case the boundaries appear in the time direction, which is described by the boundary state. The Hamiltonian is same as the one without a boundary $H$. Alternatively, one could take the $x$-direction as the Euclidean time. In this case the Hamiltonian is no longer same as the one without boundary, but depends on the two boundary conditions, denoted by $H_B$\cite{Ghoshal:1993tm}. The two-point correlation function can be expressed as
\begin{eqnarray}\label{twopointviersion2}
\langle O(0,0)O(L,0)\rangle_{\text{strip}} &=&~_B\!\Bra{0}\mathcal{T}_{x}O(0,0)O(L,0)\Ket{0}_B\nonumber \\
&=&~_B\!\Bra{0}O(0,0)e^{-L H_B}O(0,0)\Ket{0}_B \nonumber \\
&=&\sum_n e^{-L (E_n-E_0)}~_B\!\Bra{0}O(0,0)\ket{n}_B~_B\!\bra{n}O(0,0)\Ket{0}_B,
\end{eqnarray}
where $\mathcal{T}_{x}$ means $x$-ordering, $\ket{n}_B$ is n-th eigenstate of $H_B$. So when $L$ is large, the term that dominates the sum is associated with the first excited state $\ket{1}_B$. By comparing the exponential decay term of (\ref{twopointversion1}) with (\ref{twopointviersion2}), one could find the energy gap between the first excited state and the ground state,
\be\label{energygap}
\delta E\equiv E_1-E_0= \frac{\pi\eta_{||}}{2\tau_0},
\ee
which is related to the width of the strip and operator contents of the theory.\\

In paper \cite{Fredenhagen}  Fredenhagen shows the energy gap implies exponential decay of spatial correlations between bounded operators. For the boundary state we could establish a similar cluster theorem.\\
In general  one could introduce a norm $\Vert \cdot \Vert$ of operators acting on a Hilbert space, which is a map from a operator to a real number satisfying certain constraints \cite{Bratteli}. A bounded operator $A$ is  the one whose norm $\Vert A \Vert$ is finite. The bounded operators in QFT usually constitute a certain algebra, so-called $C^*$-algebra\cite{Haag2}. Similarly, one could define local $C^*$-algebras associated with an open spacetime region $\mathcal{O}$, denoted by $\mathcal{A}(\mathcal{O})$, for which $A\in \mathcal{A}(\mathcal{O})$ is vanishing outside of region $\mathcal{O}$. In the 2D CFT the $C^*$-algebra $\mathcal{A}(\mathcal{O})$  could be constructed by the (qusai-)primary operators with some smearing function whose suppose is in region $\mathcal{O}$\cite{Carpi:2015fga}. Specially as shown in paper \cite{Constantinescu}, the smeared chiral  vertex operators of 2D scalar field $\int f(z)e^{-i\alpha \phi(z)}dz$ is a bounded operator. In paper \cite{Carpi:2015fga} the authors discuss how to construct the local observables by the vertex algebras. Here we won't discuss the details of the construction, but only assume the local bounded operators exists, they can be constructed by (qusai-)primary operators and suitable smearing functions. \\

In 2D CFT the local obervables are associated with the intervals. Set $\mathcal{O}_1$ to be the interval $[x_1,x_2]$, the corresponding local $C^*$-algebra to be $\mathcal{A}(\mathcal{O}_1)$. Similarly, take $\mathcal{O}_2$ to be  the interval $[x_3,x_4]$ ($x_3>x_2$), its local $C^*$-algebra is $\mathcal{A}(\mathcal{O}_2)$. Asumme $x_3-x_2=L$. We would like to consider the cluster property for bounded operators  $A_1 \in \mathcal{O}_1$ and $A_2 \in \mathcal{O}_2$ in the regularized boundary state (\ref{regularizedstate}),i.e., to estimate
\be
C_{12}\equiv ~_a^\t0\!\bra{B}A_1 A_2\ket{B}^{\tau_0}_{a}- ~_a^\t0\!\bra{B}A_1\ket{B}^{\tau_0}_{a}~_a^\t0\!\bra{B} A_2\ket{B}^{\tau_0}_{a}.
\ee
It is sufficient to consider the case $~_a^\t0\!\bra{B}A_{1(2)}\ket{B}^{\tau_0}_{a}=0$. Otherwise, we could instead use the operators $A'_{1(2)}\equiv A_{1(2)}-~_a^\t0\!\bra{B}A_{1(2)}\ket{B}^{\tau_0}_{a}$ $I$, which gives $~_a^\t0\!\bra{B}A'_{1(2)}\ket{B}^{\tau_0}_{a}=0$.
The correlator $~_a^\t0\!\bra{B}A_1 A_2\ket{B}^{\tau_0}_{a}$ can be taken as correlator in a strip of width $2\tau_0$, which could be expressed as
\be
\langle A_1 A_2 \rangle_{\text{strip}}=~_B\!\Bra{0}\mathcal{T}_x A_1 A_2\ket{0}_B.
\ee
 For the algebras $\mathcal{A}(\mathcal{O})$ associated with region $\mathcal{O}$, we have the following translation property
\be
e^{-l H_B}\mathcal{A}(\mathcal{O})e^{l H_B}=\mathcal{A}({\mathcal{O}+l}),
\ee
where $H_B$ is the generator of ``time'' evolution, $O+l$ means the image of $O$ under the ``time'' translation, with $\mathcal{O}$ shifts $l$ in the $x$ direction. Let's define a region $\mathcal{O}_3$ to be
\be\label{3to1}
\mathcal{O}_3\equiv e^{L H_B}\mathcal{O}_2 e^{-L H_B}.
\ee
The overlap between $\mathcal{O}_3$ and $\mathcal{O}_{1}$ is just one point $x_3$.
The $C^*$-algebra $\mathcal{A}(\mathcal{O}_3)$ is isomorphic to $\mathcal{A}(\mathcal{O}_2)$, since the relation (\ref{3to1}). Thus for any operator $A_2\in \mathcal{A}(\mathcal{O}_2)$ there exists an operator $A_3 \in \mathcal{A}(\mathcal{O}_3)$, such that $A_2=e^{-L H_B}A_3 e^{L H_B}$, where $A_3$ is also a bounded operator. We have
\begin{eqnarray}
&&|~_B\!\Bra{0}\mathcal{T}_x A_1 A_2 \ket{0}_B|=|~_B\!\Bra{0}\mathcal{T}_x A_1e^{-LH_B} A_3 \ket{0}_B|\\ \nonumber
&&=\sum_n e^{-(E_n-E_0)L}|~_B\!\Bra{0} A_1\ket{n}_B~_B\!\bra{n} A_3 \ket{0}_B|\\ \nonumber
&&\le e^{-\delta E L}|~_B\!\Bra{0}\mathcal{T}_x A_1 A_3 \ket{0}_B|
\end{eqnarray}
For a bounded operator $A_3\in \mathcal{A}(\mathcal{O}_3)$, we can estimate
\be\label{Cauchy}
|~_B\!\Bra{0}\mathcal{T}_x A_1 A_3  \ket{0}_B|=|~_B\!\Bra{0}\mathcal{T}_x( A_1) \mathcal{T}_x (A_3) \ket{0}_B|\le  \Vert (\mathcal{T}_x A_1)^\dagger \ket{0}_B\Vert\cdot \Vert \mathcal{T}_x A_3 \ket{0}_B\Vert,
\ee
where we have used the fact that  $\mathcal{O}_3$ and $\mathcal{O}_2$ have no overlap and  Cauchy-Schwarz inequality. Note that if $\mathcal{O}_3$ and $\mathcal{O}_2$ have overlaps the equality may break down. Since $A_1$ and $A_3$ are both bounded operator, we expect the last term in (\ref{Cauchy}) is bounded by some constant $M_{12}$.\\
Thus we could estimate
\be \label{estimation}
~_a^\t0\!\bra{B}A_1 A_2\ket{B}^{\tau_0}_{a}\le e^{-\frac{L\pi\eta_{||}}{2\tau_0}}M_{12},
\ee
$M_{12}$ is finite. For $~_a^\t0\!\bra{B}A_{1(2)}\ket{B}^{\tau_0}_{a}\ne 0$  taking $A'_{1(2)}$ into (\ref{estimation}), we will obtain a similar estimation. Therefore,
\be\label{cluster}
C_{12}\le e^{-\frac{L\pi\eta_{||}}{2\tau_0}}M_{12}.
\ee

\subsection{Bell inequality as witness to detect entanglement}
Quantum entanglement of a given state in QFT is directly related to the operators correlation strength in this state. This is not clear when we use entanglement entropy(EE) to describe the entanglement property of a state, since EE deals with the state itself, unrelated to the operators. On the other hand Bell inequalities are based on the operators and their correlations in a state. In general violation of Bell inequalities means existence of quantum entanglement. Let's recall some basic definitions on Bell inequalities.\\
The general set-up can be found in paper \cite{Summer1}\cite{Summer2}. In QFT we consider a state $\omega$ on a bipartite system $\mathcal{O}_1$ and $\mathcal{O}_2$, where $\mathcal{O}_1$ and $\mathcal{O}_2$ are spacelike.  In 2D CFT one could assume they are two intervals with  distance $L$. For Hermitean operators $A_1,A'_1\in \mathcal{A}(\mathcal{O}_1)$ and $A_2,A'_2\in \mathcal{A}(\mathcal{O}_2)$ we define a quantity
\be
\gamma(\mathcal{O}_1, \mathcal{O}_2) =\frac{1}{2}|\omega(A_1(A_2+A'_2)+A'_1(A_2-A'_2))|.
\ee
 State $\omega$ is said to satisfy  the  Bell inequality  of CHSH form if
\be
\gamma(\mathcal{O}_1, \mathcal{O}_2)\le 1,
\ee
for all the hermintean operators $A_1,A'_1,A_2,A'_2$ whose norm is bounded by 1. If  existing operators such that $\gamma>1$, there is quantum entanglement between $\mathcal{O}_1$ and $\mathcal{O}_2$ in state $\omega$. \\
We would like to estimate the CHSH-Bell inequality for the regularized boundary state (\ref{regularizedstate}). Assume $\omega(\cdot)=~_a^\t0\!\bra{B}\cdot \ket{B}^{\tau_0}_{a}$, by using the cluster estimation (\ref{cluster}) we have
\begin{eqnarray}
&&|\omega(A_1(A_2+A'_2)+A'_1(A_2-A'_2))|\nonumber \\
&&\quad \le |\omega(A_1)(\omega(A_2)+\omega(A'_2))+\omega(A'_1)(\omega(A_2)-\omega(A'_2))|+ 2M e^{-\frac{L\pi\eta_{||}}{2\tau_0}}\nonumber \\
&&\quad \le 2+2M e^{-\frac{L\pi\eta_{||}}{2\tau_0}},
\end{eqnarray}
where $M$ is a finite number related to the cluster property, one could obtain the last step by using,
\begin{eqnarray}
&&\omega(A_1)(\omega(A_2)+\omega(A'_2))+\omega(A'_1)(\omega(A_2)-\omega(A'_2))\nonumber \\
&&=\frac{1}{2}\omega(1+A_1)\omega(1+A_2)\omega(A'_1)+\frac{1}{2}\omega(1+A_1)\omega(1-A_2)\omega(A'_2)\nonumber \\
&&-\frac{1}{2}\omega(1-A_1)\omega(1+A_2)\omega(A'_1)-\frac{1}{2}\omega(1-A_1)\omega(1-A_2)\omega(A'_1)\nonumber \\
&&\le \frac{1}{2}\omega(1+A_1)\omega(1+A_2)+\frac{1}{2}\omega(1+A_1)\omega(1-A_2)\nonumber \\
&&+\frac{1}{2}\omega(1-A_1)\omega(1+A_2)+\frac{1}{2}\omega(1-A_1)\omega(1-A_2)\nonumber \\
&&=2
\end{eqnarray}
Therefore we obtain
\be\label{estimationBell}
\gamma(\mathcal{O}_1, \mathcal{O}_2) \le 1+ M  e^{-\frac{L\pi\eta_{||}}{2\tau_0}}.
\ee
 One could see that if $L\gg \tau_0$, $\gamma \to 1$, which means quantum entanglement is vanishing in this limit. This is consistent with the result that boundary states has no real space entanglement \cite{Miyaji:2014mca}, since $\tau_0\to \epsilon$ ($\epsilon$ is the UV cut-off of the theory) leads to $\gamma\to 1$.\\
The vacuum state of CFT is expected to have quantum entanglement even if two regions are far away. This is due to the scale invariance of vacuum state. As the theorem 4.4 in paper \cite{Summer1}  shows for any two spacelike wedges $W_1$ and $W_2$, $\gamma(W_1, W_2)=\gamma(W'_i,W_i)$ ($i=1,2$, $W'$ is the complementary of $W$) by using the scaling invariance of the vacuum state. In  paper \cite{Summer2} the authors prove maximal violation, i.e., $\gamma(W'_i,W_i)=\sqrt{2}$.
But the regularized boundary state is not scale invariant any more as we have mentioned. For a theory with a mass gap $m$ the estimation of $\gamma(\mathcal{O}_1, \mathcal{O}_2)$ is similar as (\ref{estimationBell}) with $\tau_0\to 1/m$ \cite{Summer1}.  This can be seen as an evidence that the regularized boundary state can be associated with the vacuum state of a non-conformal field theory, which has a mass scale $m$\cite{Miyaji:2014mca}\cite{Cardy:2017ufe}.\\
~\\
The real spacetime  entanglement  in boundary state of 2D CFT is vanishing. This conclusion is closely related to the exponential decay behavior of correlation function. But as we can see in (\ref{poly}) the correlation function is not exponentially decaying in the Neumann boundary state in 4D. As argued in paper \cite{Miyaji:2014mca}\cite{Cardy:2017ufe} the boundary state can be associated with the vacuum state of the Hamiltonian of a new theory by a relevant deformation of the original CFT,
\begin{eqnarray}
 H_M=H_{CFT}+M^{2-\Delta} \int dx O(x),
 \end{eqnarray}
where $M$ is the mass scale of this massive deformation, $\Delta$ is the conformal dimension of the operator $O$. For example the Dirichlet boundary state $\ket{B}_{-}$ can be seen as the vacuum state of the massive free scalar theory with
\begin{eqnarray}
H=\int d^3x (\pi(x)^2+M \phi(x)^2),
\end{eqnarray}
with $M\sim 1/\epsilon$. The theory will flow into a trivial IR theory which has no propagating degrees of freedom, thus no real space entanglement \cite{Miyaji:2014mca}.  The correlator in Dirichlet boundary state is indeed exponentially decaying\footnote{One could also check the correlator
 \begin{eqnarray}
~_{-}^{\tau_0}\!\bra{B}\pi(x,0)\pi(y,0)\ket{B}^{\tau_0}_{-}\sim  \tau_0 ^{-3}coth\left[\frac{\pi  r}{2 \tau_0 }\right] cosh\left[\frac{\pi  r}{2 \tau_0 }\right]^2\sim e^{-\frac{\pi r}{\tau_0}}.
\end{eqnarray}} (3.16).
But the Neumann boundary state seems not like this. It is probably that the real space entanglement is not vanishing in Neumann boundary state\footnote{In paper \cite{Mollabashi:2013lya} the authors also notice that in the Direchlet boundary state the degrees of freedom decrease , but their results suggest the Neumann boundary condition increases the IR degrees of freedom.}. The polynomial decay of correlator  is a hint on this.

\section{Time evolution of boundary state}\label{timeevolution}
As we have mentioned the boundary state is not time-independent. In paper \cite{Calabrese:2006rx}\cite{Calabrese:2007rg} the authors suggested the boundary state can be associated with a quantum quench process, finally the system will locally approach to a thermal state, even though the whole system remains a pure state. Later the result is generalized to more general boundary state in paper \cite{Cardy:2015xaa}. We have many general characterizations to describe the equilibrium phenomena. One of them is to investigate the Kubo-Martin-Schwinger (KMS) condition. It concerns with the correlation relation of operators in the theory, and emphasizes the algebraic properties of the observables. We will first introduce the KMS condition in field theory, and check  whether the regularized boundary states(\ref{regularizedstate}) after long time evolution satisfy such condition.

\subsection{KMS condition for field theory}\label{KMSconditon}

Assume the Hamiltonian of the field theory is $H$, the corresponding one-parameter group is
\be
\tau_z (A)\equiv e^{i z H}A e^{-i z H},
\ee
where $A$ is the operator in the Hilbert space of theory, $z$ is complex constant. A state $\omega$ in quantum field theory can be taken as positive linear functional of operators, such that $\omega(A^\dagger A)\ge 0$ for any operator $A$, and $\omega(1)=1$. A state $\omega$ satisfies the KMS condition, if
\be\label{KMS}
\omega(A \tau_{i\beta}(B))=\omega(B A),
\ee
for any operator $A, B$, where $\beta$ is a real constant. At the same time we ask the function $F(z)\equiv \omega(A\tau_{z}(B))$ is analytic in the region $0<Im(z)<\beta$(if $\beta>0$).\\
 Such state is called a (global ) $(\tau,\beta)$-KMS state. $\beta=0$ is a special case, $\omega$ is a trace state, $\omega(AB)=\omega(BA)$ for any $A,B$. In this sense the KMS condition reflects the deviation of $\omega$ from being a trace state. \\
KMS states  are usually related to thermal equilibrium, so the state is not expected to be time-independent. The following proposition \cite{Robinson} ensures the KMS state is invariant under time evolution.

\emph{If $\omega$ is a $(\tau,\beta)$-KMS state, with $\beta \ne 0$, it follows that $\omega$ is time-invariant, i.e.,
\be
\omega(\tau_t (A))=\omega(A),
\ee
for all operators $A$ and $t\in \mathbb{R}$.
}\\

We show the proof of the theorem in the appendix \ref{theorem}.
So a necessary condition for a state being a KMS state is that the state must be time-invariant. In the following we will discuss the time evolution of the regularized boundary state, i.e., $\ket{\psi(t)}_a\equiv e^{-i H t}\ket{B}^{\tau_0}_a$. Consider the state
\be
\ket{\psi(\infty)}\equiv \lim_{t\to \infty} e^{-i H t}\ket{B}^{\tau_0}_a,
\ee
it is indeed time-independent, since $e^{-iH t_0}\ket{\psi(\infty)}=\lim_{t\to +\infty} e^{-i H (t-t_0)}\ket{B}^{\tau_0}_a=\ket{\psi(\infty)}$.\\
The KMS condition can be expressed in diverse ways, (\ref{KMS}) is the most convenient for our discussion.

\subsection{Time evolution of boundary state and KMS state}\label{KMSsection}
We will consider the state $\ket{\psi(t)}_a\equiv e^{i H t}\ket{B}^{\tau_0}_a$, and analysis the correlator in such state.
\subsubsection{Free field theory}\label{freescalarKMS}
Let's first see the free scalar field. As we have mentioned in section \ref{regboundarysection} the correlation function of $a_{\bm{k}}$ and $a^\dagger_{\bm{k}}$ (\ref{cacorrelation}) implies some information is hidden in the regularized boundary state. Consider the two-point function of $\phi(x,t_x)$ and $\phi(x,t_y)$,
\begin{eqnarray}
~_{\pm}\!\bra{\psi(t)}\phi(x,t_x)\phi(y,t_y)\ket{\psi(t)}_{\pm}=~_{\pm}^{\tau_0}\!\bra{B}\phi(x,t_x+t)\phi(x,t_y+t)\ket{B}^{\tau_0}_\pm.
\end{eqnarray}
By using (\ref{cacorrelation}) we have
\begin{eqnarray}\label{twoscalarcorrelationtime}
&&~_{\pm}\!\bra{\psi(t)}\phi(x,t_x)\phi(y,t_y)\ket{\psi(t)}_{\pm}\nonumber \\
&&=\frac{1}{(2\pi)^3}\int \frac{d^3k}{2E_k}\frac{1}{e^{4\tau_0 E_k}-1}\Big(e^{4\tau_0E_k}e^{iE_k(t_y-t_x)}+e^{-iE_k(t_y-t_x)}\Big)e^{i\bm{k}\cdot (\bm{x-y})}\nonumber \\
&&\pm\frac{1}{\sqrt{2\pi}r}\int_{-\infty}^{+\infty}dk \Big(e^{ik(t_x+t_y)}\sin(k r)\frac{e^{2\tau_0 k}}{e^{4\tau_0 k}-1}\Big)e^{2ikt} dk,
\end{eqnarray}
where $r$ is the distance $|\bm{x-y}|$.
The first line of the result is time-independent. The second line depends on time, but the integration is  the form $\int dk f(k)e^{i k t}$, $f(k)$ is smooth and exponential decay when $|k|$ is large, so according to Riemann-Lebesgue lemma, the integral is vanishing in the limit $t\to+\infty$. Thus we obtain
\begin{eqnarray}\label{thermal4D}
&&~_{\pm}\!\bra{\psi(+\infty)}\phi(x,t_x)\phi(y,t_y)\ket{\psi(\infty)}_{\pm}\nonumber \\
&&=\frac{1}{(2\pi)^3}\int \frac{d^3k}{2E_k}\frac{1}{e^{4\tau_0 E_k}-1}\Big(e^{4\tau_0E_k}e^{iE_k(t_y-t_x)}+e^{-iE_k(t_y-t_x)}\Big)e^{i\bm{k}\cdot (\bm{x-y})}.
\end{eqnarray}
Finally one could check
\begin{eqnarray}
&&~_{\pm}\!\bra{\psi(+\infty)}\phi(x,t_x)\tau_{i \beta}[\phi(y,t_y)]\ket{\psi(\infty)}_{\pm}|_{\beta=4\tau_0}\nonumber \\
&&=\frac{1}{(2\pi)^3}\int \frac{d^3k}{2E_k}\frac{1}{e^{4\tau_0 E_k}-1}\Big(e^{iE_k(t_y-t_x)}+e^{4\tau_0E_k}e^{-iE_k(t_y-t_x)}\Big)e^{i\bm{k}\cdot (\bm{x-y})}\nonumber \\
&&=~_{\pm}\!\bra{\psi(+\infty)}\phi(y,t_y)\phi(x,t_x)\ket{\psi(\infty)}_{\pm}.
\end{eqnarray}
We arrive at the conclusion that the two-point correlation function satisfies the KMS condition with $\beta=4\tau_0$ in the limit $t\to+\infty$. Also notice that the final KMS state does not depend on the initial boundary condition($\pm$). \\
The higher order correlation functions are not so easy to obtain. In free field theory any operator can be constructed by $a_{\bm{k}}$ and $a^\dagger_{\bm{k}}$. From (\ref{cacorrelation}) we have
\be
~_{\pm}^\t0\!\bra{B}a_{\bm k}\tau_{i\beta}(a^\dagger_{\bm p})\ket{B}^\t0_{\pm}|_{\beta=4\tau_0}=~_{\pm}^\t0\!\bra{B}a_{\bm k}a^\dagger_{\bm p}\ket{B}^\t0_{\pm}e^{-4\tau_0 E_k}=~_{\pm}^\t0\!\bra{B}a^\dagger_{\bm p} a_{\bm k}\ket{B}^\t0_{\pm}|_{\beta=4\tau_0}.
\ee
This term is time-independent, other combinations $a_{\bm{k}}a_{\bm{p}}$ or $a^\dagger_{\bm{k}}a^\dagger_{\bm{p}}$ depends on time and vanish in the limit $t\to \infty$. This is the reason why the two-point correlation function satisfies the KMS-condition. For higher order correlation functions we have to calculate the correlation such as $~_{\pm}^\t0\!\bra{B}a_{\bm k_1}...a^\dagger_{\bm p_1}...\ket{B}^\t0_{\pm}$. Let's consider the four-point function. The typical one is $~_{\pm}^\t0\!\bra{B}a^\dagger_{\bm{p}_2}a^\dagger_{\bm{p}_1}a_{\bm{k}_1}a_{\bm{k}_2}\ket{B}^{\tau_0}_{\pm}$. Other ones can be translated to it by commutation relation. In the Appendix \ref{4point} we give the detail of the calculation on this four-point function. The result is
\begin{eqnarray}
&&~_{\pm}^\t0\!\bra{B}a^\dagger_{\bm{p}_2}a^\dagger_{\bm{p}_1}a_{\bm{k}_1}a_{\bm{k}_2}\ket{B}^{\tau_0}_{\pm}\nonumber \\
&&=\delta(\bm{p_1}+\bm{p_2})\delta(\bm{k_1}+\bm{k_2}) \frac{e^{2\tau_0(E_{k_2}+E_{p_2})}}{(e^{4\tau_0E_{p_2}}-1)(e^{4\tau_0E_{k_2}}-1)}\nonumber \\
&&+\Big(\delta(\bm{p}_1-\bm{k}_1)\delta(\bm{p}_2-\bm{k}_2)+\delta(\bm{p}_1-\bm{k}_2)\delta(\bm{p}_2-\bm{k}_1)\Big) \frac{1}{(e^{4\tau_0E_{p_1}}-1)(e^{4\tau_0E_{p_2}}-1)}.
\end{eqnarray}
Consider the time evolution we obtain
\begin{eqnarray}
&&~_{\pm}\!\bra{\psi(t)}a^\dagger_{\bm{p}_2}a^\dagger_{\bm{p}_1}a_{\bm{k}_1}a_{\bm{k}_2}\ket{\psi(t)}_{\pm}\nonumber \\
&&=\delta(\bm{p_1}+\bm{p_2})\delta(\bm{k_1}+\bm{k_2}) \frac{e^{2\tau_0(E_{k_2}+E_{p_2})}}{(e^{4\tau_0E_{p_2}}-1)(e^{4\tau_0E_{k_2}}-1)}e^{it(E_{p_1}+E_{p_2}-E_{k_1}-E_{k_2})}\nonumber \\
&&+\Big(\delta(\bm{p}_1-\bm{k}_1)\delta(\bm{p}_2-\bm{k}_2)+\delta(\bm{p}_1-\bm{k}_2)\delta(\bm{p}_2-\bm{k}_1)\Big) \frac{1}{(e^{4\tau_0E_{p_1}}-1)(e^{4\tau_0E_{p_2}}-1)},
\end{eqnarray}
where the term in the first line is time-dependent (if $E_{p_1}+E_{p_2}-E_{k_1}-E_{k_2}\ne 0$), the second line is always invariant under time evolution.
The operators in the spacetime usually are like the form,
\be
\int d\bm{p_1}d\bm{p_2}d\bm{k_1}d\bm{k_2}f(\bm{p_1,p_2,k_1,k_2})~_{\pm}\!\bra{\psi(t)}a^\dagger_{\bm{p}_2}a^\dagger_{\bm{p}_1}a_{\bm{k}_1}
a_{\bm{k}_2}\ket{\psi(t)}_{\pm}.
\ee
Assume the function $f(\bm{p_1,p_2,k_1,k_2})$ is smooth enough, in the limit $t\to \infty$ finally only the term in the second line survives. One could
check
\begin{eqnarray}
&&\int d\bm{p_1}d\bm{p_2}d\bm{k_1}d\bm{k_2}f(\bm{p_1,p_2,k_1,k_2})~_{\pm}\!\bra{\psi(+\infty)}a^\dagger_{\bm{p}_2}a^\dagger_{\bm{p}_1}a_{\bm{k}_1}
\tau_{i\beta}(a_{\bm{k}_2})\ket{\psi(+\infty)}_{\pm}|_{\beta=4\tau_0}\nonumber \\
&&=\int d\bm{p_1}d\bm{p_2}d\bm{k_1}d\bm{k_2}f(\bm{p_1,p_2,k_1,k_2})~_{\pm}\!\bra{\psi(+\infty)}a_{\bm{k}_2}a^\dagger_{\bm{p}_2}a^\dagger_{\bm{p}_1}a_{\bm{k}_1}
\ket{\psi(+\infty)}_{\pm},
\end{eqnarray}
or
\begin{eqnarray}
&&\int d\bm{p_1}d\bm{p_2}d\bm{k_1}d\bm{k_2}f(\bm{p_1,p_2,k_1,k_2})~_{\pm}\!\bra{\psi(+\infty)}a^\dagger_{\bm{p}_2}a^\dagger_{\bm{p}_1}\tau_{i\beta}(a_{\bm{k}_1}
a_{\bm{k}_2})\ket{\psi(+\infty)}_{\pm}|_{\beta=4\tau_0}\nonumber \\
&&=\int d\bm{p_1}d\bm{p_2}d\bm{k_1}d\bm{k_2}f(\bm{p_1,p_2,k_1,k_2})~_{\pm}\!\bra{\psi(+\infty)}a_{\bm{k}_1}a_{\bm{k}_2}a^\dagger_{\bm{p}_2}a^\dagger_{\bm{p}_1}
\ket{\psi(+\infty)}_{\pm},
\end{eqnarray}
and so on. The argument can be generalized to any point correlation function. In general by induction we could obtain the time-independent term of the typical $2n$-correlation
\begin{eqnarray}
&&~_{\pm}\!\bra{\psi(t)}a^\dagger_{\bm{p}_n}...a^\dagger_{\bm{p}_1}a_{\bm{k}_1}...a_{\bm{k}_n}\ket{\psi(t)}_{\pm}|_{\text{time-independent}}\\
&&=\sum_{\sigma(i_j)}\prod_{j=1,...,n}\delta(\bm{p}_{\sigma(i_j)}-\bm{k}_j)\prod_{m=1,...,n}\frac{1}{e^{4\tau_0E_{k_m}}-1},
\end{eqnarray}
where $\sigma(i_j)$ means all the permutations.

\subsubsection{2D CFT}\label{2DCFT}
Consider the time-dependent two-point function,
\be
~_{a}\!\bra{\psi(t)}O(w_1,\bar w_1)O(w_2,\bar w_2)\ket{\psi(t)}_{a}.
\ee
In the Heisenberg picture it becomes
\be
~_a^\t0\!\bra{B}O(w_1-t,\bar w_1+t)O(w_2-t,\bar w_2+t)\ket{B}^{\tau_0}_{a}.
\ee
By using (\ref{crossration}) and (\ref{2D2point}), and take the limit $t\to +\infty$, we obtain
\begin{eqnarray}\label{Thermal2point}
&&~_{a}\!\bra{\psi(+\infty)}O(w_1,\bar w_1)O(w_2,\bar w_2)\ket{\psi(+\infty)}_{a}\nonumber \\
&&=\Big(\frac{\pi}{4\tau_0}\Big)^{4h}\Big[\sinh(\frac{\pi(w_1-w_2)}{4\tau_0})\sinh(\frac{\pi(\bar w_1-\bar w_2)}{4\tau_0})\Big]^{-2h}.
\end{eqnarray}
This expression is not the correct correlation function in Minkowski spacetime. It is invariant under permutation of $O(w_1,\bar w_1)$ and $O(w_2,\bar w_2)$.
We need to make clear the correlation function in Minkowski spacetime before checking the KMS relation. \\
In general the Euclidean correlation function
$\langle O(w_1,\bar w_1)O(w_2,\bar w_2)...O(w_n,\bar w_n)\rangle_E$ is invariant under permutation of the position. But the order of positions in Minkowski spacetime is very important. The correlation functions in Minkowski spacetime can be obtained upon analytic continuation of Euclidean correlators. A way to realize this process is called $i\epsilon$ prescription \cite{Hartman:2015lfa}. For some ordering Minkowski correlation function could be obtained by analytically continuing $\tau^E_i\to it_i+\epsilon$ (or equally  $w^E_i\to w_i+i \epsilon_i$ and $\bar w^E_i\to \bar w_i-i \epsilon_i$) with $\epsilon_i$ ordered, more precisely,
\begin{eqnarray}
&&\langle O(w_1,\bar w_1)O(w_2,\bar w_2)...O(w_n,\bar w_n)\rangle_M\\ \nonumber
&&=\lim_{\epsilon_i\to 0} \langle O(w_1+i \epsilon_1,\bar w_1-i \epsilon_1)O(w_2+i\epsilon_2,\bar w_2-i\epsilon_2)...O(w_n+i\epsilon_n,\bar w_n-i\epsilon_n)\rangle_E,\\
&&\text{with }\quad \epsilon_1>\epsilon_2>...>\epsilon_n \nonumber
\end{eqnarray}
With this we could obtain the correct Minkowski correlator,
\begin{eqnarray}\label{}
&&~_{a}\!\bra{\psi(+\infty)}O(w_1,\bar w_1)O(w_2,\bar w_2)\ket{\psi(+\infty)}_{a}\nonumber \\
&&=\Big(\frac{\pi}{4\tau_0}\Big)^{4h}\Big[\sinh(\frac{\pi(w_1-w_2+i\epsilon)}{4\tau_0})\sinh(\frac{\pi(\bar w_1-\bar w_2-i\epsilon)}{4\tau_0})\Big]^{-2h},
\end{eqnarray}
and
\begin{eqnarray}
&&~_{a}\!\bra{\psi(+\infty)}O(w_2,\bar w_2)O(w_1,\bar w_1)\ket{\psi(+\infty)}_{a}\nonumber \\
&&=\Big(\frac{\pi}{4\tau_0}\Big)^{4h}\Big[\sinh(\frac{\pi(w_1-w_2-i\epsilon)}{4\tau_0})\sinh(\frac{\pi(\bar w_1-\bar w_2+i\epsilon)}{4\tau_0})\Big]^{-2h},
\end{eqnarray}
where $\epsilon\equiv\epsilon_1-\epsilon_2>0$. Notice the sign difference before $\epsilon$, which is quite important for keeping causal relation in Minkowskin spacetime. If $O(w_1,\bar w_1)$ and $O(w_2,\bar w_2)$ are timelike, e.g., in the region $w_1<w_2$ and $\bar w_1>\bar w_2$, we have
\begin{eqnarray}
&&~_{a}\!\bra{\psi(+\infty)}O(w_1,\bar w_1)O(w_2,\bar w_2)\ket{\psi(+\infty)}_{a}\nonumber \\
&&=e^{2i\pi h}\Big(\frac{\pi}{4\tau_0}\Big)^{4h}\Big[\sinh(\frac{\pi(w_1-w_2)}{4\tau_0})\sinh(\frac{\pi(\bar w_1-\bar w_2)}{4\tau_0})\Big]^{-2h},
\end{eqnarray}
while
\begin{eqnarray}\label{time2point}
&&~_{a}\!\bra{\psi(+\infty)}O(w_2,\bar w_2)O(w_1,\bar w_1)\ket{\psi(+\infty)}_{a}\nonumber \\
&&=e^{-2i\pi h}\Big(\frac{\pi}{4\tau_0}\Big)^{4h}\Big[\sinh(\frac{\pi(w_1-w_2)}{4\tau_0})\sinh(\frac{\pi(\bar w_1-\bar w_2)}{4\tau_0})\Big]^{-2h}.
\end{eqnarray}
 If they are spacelike, e.g., in the region $w_1>w_2$ and $\bar w_1>\bar w_2$, one could show they do commutate.\\
  Now let's see the KMS condition. Consider $w_1<w_2$ and $\bar w_1>\bar w_2$,
 \begin{eqnarray}
 &&~_{a}\!\bra{\psi(+\infty)}O(w_1,\bar w_1)\tau_{i\beta}(O(w_2,\bar w_2))\ket{\psi(+\infty)}_{a}|_{\beta=4\tau_0}\\ \nonumber
 &&=\Big(\frac{\pi}{4\tau_0}\Big)^{4h}\Big[\sinh(\frac{\pi(w_1-w_2+i\beta+i\epsilon)}{4\tau_0})\sinh(\frac{\pi(\bar w_1-\bar w_2-i\beta-i\epsilon)}{4\tau_0})\Big]^{-2h}|_{\beta=4\tau_0} \nonumber \\
 &&=e^{-2i\pi h}\Big(\frac{\pi}{4\tau_0}\Big)^{4h}\Big[\sinh(\frac{\pi(w_1-w_2)}{4\tau_0})\sinh(\frac{\pi(\bar w_1-\bar w_2)}{4\tau_0})\Big]^{-2h},
 \end{eqnarray}
 where the phase is from the anti-holomorphic part. The result is same as (\ref{time2point}).  We should not be surprise about this result, since (\ref{Thermal2point}) is the exact thermal two-point correlation function in 2D CFT, which can be obtained by conformal mapping from complex plane to a cylinder.\\
 To obtain $n$-point function in a strip we need to know the corresponding $n$-point function on UHP, which can be associated with $2n$-point function on the whole complex plane. Unfortunately, we still can't gain the result for general theory even in the limit $t\to \infty$. We will only consider 2D free scalar. In principle the argument in section \ref{freescalarKMS} can also
be used for 2D free theory. But here we do not use the Fock space formulism, only take advantage of the conformal symmetry and ``image method'' to calculate the correlation functions on UHP. \\

It is pointed out by Cardy in paper \cite{Cardy2} that the $n$-point functions on UHP can be associated with $2n$-point functions on the entire plane, which are regard as functions of $2n$ holomorphic variables $z_1,z_2...,z_{2n}$ with $z_{n+i}=z_i^*$. Thus the $n$-point functions are replaced by the holomorphic part of $2n$-point functions, the boundary conditions are used to determine the solutions of the differential equations satisfied by the $2n$-point function. This process can be heuristically written as
\be\label{3pointUHP}
\langle O(z_1,\bar z_1)...O(z_n,\bar z_n)\rangle_{\text{UHP}}=\langle O(z_1)...O(z_n)\tilde{O}(z^*_1)...\tilde{O}(z^*_n)\rangle_{\text{$z$-plane}},
\ee
where $\tilde{O}$ is the ``image'' of $O$, which is related to $O$ by a parity transformation, determined by boundary condition.
\\
Let's consider the vertex operator $O_{\alpha}\equiv e^{-i\alpha \phi}$ with conformal dimension $h=\bar h=\frac{\alpha^2}{2}$\footnote{One could consider more general operator, such as $e^{i\alpha\phi}+e^{-i\alpha \phi}$, whose conformal block is non-trivial. The correlation function on UHP of this operator can be obtained by relating it to $e^{i\alpha \phi}$\cite{Guo:2015uwa}  }. For Neumann(+) and Dirichlet($-$) boundary condition, the respective parity transformations are $\phi(z,\bar z)\to \eta \phi(\bar z,z)$ with $\eta=\pm 1$. Consider the 3-point function
\begin{eqnarray}
&&~_\pm^\t0\!\bra{B}O_{\alpha_1}(w_1,\bar w_1)O_{\alpha_2}(w_2,\bar w_2)O_{\alpha_3}(w_3,\bar w_3)\ket{B}^{\tau_0}_{\pm}\nonumber \\
&&=\prod_{i=1,2,3}w'(z_i)^{-h_i} \bar w' (\bar z_i)^{-h_i} \langle O_{\alpha_1}(z_1,\bar z_1)O_{\alpha_2}(z_2,\bar z_2)O_{\alpha_3}(z_3,\bar z_3)\rangle_{\text{UHP}},
\end{eqnarray}
and assume $\alpha_1+\alpha_2+\alpha_3=0$. By the ``image'' method we have
\begin{eqnarray}
&&\langle O_{\alpha_1}(z_1,\bar z_1)O_{\alpha_2}(z_2,\bar z_2)O_{\alpha_3}(z_3,\bar z_3)\rangle_{\text{UHP}}\nonumber \\
&&=\langle \prod_{j=1,...,6}e^{i\alpha_j \phi(z_j)}\rangle,
\end{eqnarray}
where $z_{3+i}\equiv z^*_i$ and $\alpha_{3+i}\equiv \pm \alpha_{i}$ (the sign is related to the boundary conditions) with $i=1,2,3$. The neutrality condition is satisfied, i.e., $\sum_{i=1,...,6}\alpha_i=0$, one obtains
\begin{eqnarray}
&&\langle O_{\alpha_1}(z_1,\bar z_1)O_{\alpha_2}(z_2,\bar z_2)O_{\alpha_3}(z_3,\bar z_3)\rangle_{\text{UHP}}=\prod_{i<j\le 6}(z_i-z_j)^{\alpha_i \alpha_j}\nonumber \\
&&=\langle \prod_{j=1,2,3}e^{i\alpha_j \phi(z_j)}\rangle \langle \prod_{j=4,5,6}e^{i\alpha_j \phi(z_j)}\rangle \prod_{i_1=1,2,3}\prod_{j_1=4,5,6}(z_{i_1}-z_{j_1})^{\alpha_{i_1}\alpha_{j_1}}.
\end{eqnarray}
Now consider the time evolution and take the limit $t\to \infty$, we have $z_{i_1}\sim e^{-t}\to 0$ and $z_{j_1}\sim e^t\to \infty$ for $i_1=1,2,3$ and $j_1=4,5,6$, which lead to $\prod_{i_1=1,2,3}\prod_{j_1=4,5,6}|z_{i_1}-z_{j_1}|^{\alpha_{i_1}\alpha_{j_1}}\to \prod_{i_1=1,2,3}\prod_{j_1=4,5,6}|z_{j_1}|^{2\alpha_{i_1}\alpha_{j_1}}=1$. Therefore
\be
\langle O_{\alpha_1}(z_1,\bar z_1)O_{\alpha_2}(z_2,\bar z_2)O_{\alpha_3}(z_3,\bar z_3)\rangle_{\text{UHP}}=\langle \prod_{j=1,2,3}e^{i\alpha_j \phi(z_j)}\rangle \langle \prod_{j=4,5,6}e^{i\alpha_j \phi(z_j)}\rangle.
\ee
Finally we have
\begin{eqnarray}\label{thermal3point}
&&~_\pm\!\bra{\psi(+\infty)}O_{\alpha_1}(w_1,\bar w_1)O_{\alpha_2}(w_2,\bar w_2)O_{\alpha_3}(w_3,\bar w_3)\ket{\psi(+\infty)}_{\pm}\nonumber \\
&&=\frac{\Big(\frac{\pi}{4\tau_0}\Big)^{m+n+l}}{\sinh^m \frac{\pi(w_1-w_2)}{4\tau_0}\sinh^n \frac{\pi(w_1-w_3)}{4\tau_0}\sinh^l \frac{\pi(w_2-w_3)}{4\tau_0}\sinh^m \frac{\pi(\bar w_1-\bar w_2)}{4\tau_0}\sinh^n \frac{\pi(\bar w_1-\bar w_3)}{4\tau_0}\sinh^l \frac{\pi(\bar w_2-\bar w_3)}{4\tau_0}},\nonumber \\
~
\end{eqnarray}
where $m=h_1+h_2-h_3$, $n=h_1+h_3-h_2$, $l=h_2+h_3-h_1$. The result is the exact  thermal 3-point in 2D CFT, with temperature $T=1/4\tau_0$. It is straightforward to derive $n$-point function by replacing $3$ with $n$. \\
For general 3-point correlation function we could argue it should be this form.
By using (\ref{3pointUHP}), in the limit $t\to \infty$, $z_i$ will be separated from its image $z^*_i$,  the 6-point correlation function should satisfy the clustering property, i.e.,
\be
\lim_{t\to \infty}\langle O(z_1,\bar z_1)...O(z_3,\bar z_3)\rangle_{\text{UHP}}\simeq\langle O(z_1)...O(z_3)\rangle \langle\tilde{O}(z^*_1)...\tilde{O}(z^*_3)\rangle_{\text{$z$-plane}}.
\ee
The form of 3-point correlation function on $z$-plane is universal up to a coupling constant. Thus the final result should be same as (\ref{thermal3point}). This argument breaks down for $n\ge 4$, since the 4-point function depends on the details of the theory, its operator content and their fusion properties.

\section{General analysis}\label{general}

In previous sections we discuss the entanglement properties of the regularized boundary states as well as its time evolution. We have two interesting results:(1) Exponential decay of correlation in the initial state. (2) Correlation function of local operators satisfy KMS condition. It is not clear whether these two results have some relations, but they are both associated with the parameter $\tau_0$.

\subsection{General property from KMS condition}\label{generalanalysis}
We have shown in the limit $t\to \infty$, the 2-point correlation function (\ref{Thermal2point}) satisfies the KMS condition.
The 2-point function at $t=0$ (\ref{2D2point}) can be rewritten as
\be
~_a^\t0\!\bra{B}O(w_1,\bar w_1)O(w_2,\bar w_2)\ket{B}^{\tau_0}_{a}=x^{-2h}F(x)~_{a}\!\bra{\psi(+\infty)}O(w_1,\bar w_1)O(w_2,\bar w_2)\ket{\psi(+\infty)}_{a}.
\ee
In fact at $t=0$ the regularized boundary state also satisfies
\be
~_a^\t0\!\bra{B}O(w_1,\bar w_1)\tau_{i\beta}(O(w_2,\bar w_2))\ket{B}^{\tau_0}_{a}|_{\beta=4\tau_0}=~_a^\t0\!\bra{B}O(w_2,\bar w_2)O(w_1,\bar w_1)\ket{B}^{\tau_0}_{a},
\ee
because the cross ration $x$ is invariant under $\tau_{i\beta}|_{\beta=4\tau_0}$ translation. This implies the regularized boundary state has some relation with the thermal state. But it seems inconsistent with the theorem in section (\ref{KMSconditon}), which states that the KMS-condition would imply time-invariance of the state. It is obvious the boundary state is not time-independent. Actually they are consistent since the KMS-condition (\ref{KMS}) also asks the function $F(z)$ is analytic in the region $0<Im(z)<\beta$.\\
Let's see the one-point function (\ref{onepoint}) at $t=0$,
\be
~_a^\t0\!\bra{B}O(w,\bar w)\ket{B}^{\tau_0}_{a}=A^O_a \Big(\frac{\pi}{4\tau_0} \frac{1}{\cosh[(w-\bar w)\pi/4\tau_0]}\Big)^{2h}.
\ee
It satisfies $~_a^\t0\!\bra{B}\tau_{i\beta}(O(w,\bar w))\ket{B}^{\tau_0}_{a}|_{\beta=4\tau_0}=~_a^\t0\!\bra{B}O(w,\bar w)\ket{B}^{\tau_0}_{a}$, which can be seen as the condition (\ref{KMS}) with $A=I, B=O(w_1,\bar w_1)$. But the function $F_1(z)\equiv~_a^\t0\!\bra{B}\tau_{z}(O(w,\bar w))\ket{B}^{\tau_0}_{a}$ has poles in the region $0<Im(z)<4\tau_0$. To see this denote $z=z_1+i z_2$, we have
\be
F_1(z)= A^O_a\Big(\frac{\pi }{4\tau_0}\Big)^{2h} \Big(\cosh\pi (\frac{w-\bar w -2 z_1-2 i z_2}{4\tau_0})\Big)^{-2h}.
\ee
At the point $z=(w-\bar w)/2+2 i \tau_0  $, $F(z)$ will be non-analytic.  Similarly, the function $F_2(z)\equiv ~_a^\t0\!\bra{B}\tau_{z}[O(w_1,\bar w_1)O(w_2,\bar w_2)]\ket{B}^{\tau_0}_{a}$, it will also be non-analytic on the region $0<Im(z)<4\tau_0$. \\
Consider time evolution, the function $F_1(z;t)\equiv~_a\!\bra{\psi(t)}\tau_{z}(O(w,\bar w))\ket{\psi(t)}_{a}$ will have poles at point $z=(w-\bar w-2t)/2+2 i \tau_0$. In the limit $t\to \infty $ the pole will approach to infinity, thus the function $F_1(z;t)$ will be analytic on the region $0<Im(z)<4\tau_0$. Therefore the Liouville's theorem will be available in this region, which leads to the result that $\lim_{t\to \infty} F_1(z;t)\to \text{constant}$\footnote{Liouville's theorem is important for the proof of theorem in section \ref{KMSconditon}, see the Appendix \ref{theorem} for the proof. }. The effect of time evolution is to move the pole of $F_{1}(z;t)$  to infinity and make it analytic on the region $0<Im(z)<4\tau_0$.\\

As mentioned in the introduction we don't mean the state $\ket{\psi(\infty)}$ is a global KMS-state\footnote{We would like to thank Feng-Li Lin for pointing out this for me.}. The limit $t\to \infty$ is subtle, let's assume $t=T$, $T$ is a large constant. Since the system is infinite, one could always find operator $O(w,\bar w)$ such that $w-\bar w\sim T$, in this case the above argument fails. But as long as we consider $|w-\bar w|\ll T$, we find $F_1(z;T)$ will approach a constant up to a correction  $M e^{-T/\tau_0}$, where $M$ is finite. We could obtain similar result for 2-point function. Thus the final state $\ket{\psi(\infty)}$ can be only seen as a local version KMS state \cite{Buchholz:2001qj}.\\
 More precisely when saying the final state ($t\to \infty$) approach to a ($\tau,\beta$)-KMS state we means for any \emph{local} operators $A$,
\be
|\lim_{t\to\infty}\bra{\psi(t)}A\ket{\psi(t)}-\omega_{\beta}(A)|<\epsilon,
\ee
where $\epsilon$ is an arbitrary positive constant, $\omega_{\beta}$ is a global ($\tau,\beta$)-KMS state.

Generally, we could ask the following question:
start with a state $|\Psi\rangle$ in 2D CFT, which is assumed to be space-translation invariant, but not time-invariant, if the final state ($t\to \infty$) approach to a $(\tau,\beta)$-KMS state, what are the constraints on the initial state $\ket{\Psi}$?\\
Let's define the state $\ket{\Psi(\infty)}=\lim_{t\to\infty}e^{iH t}\ket{\Psi}$, where $H$ is the Hamiltonian of the theory. The KMS condition would lead to
\be
\bra{\Psi(\infty)}O(w,\bar w)\ket{\Psi(\infty)}=C_O,
\ee
where $C_O$ is a time-independent function. Since we also assume the state $\ket{\Psi}$ is space-translation invariant, $C_O$ should not depend on space coordinate, i.e., $C_O$ is a constant. When $O$ is the energy $T_{00}$, the corresponding $C'$ is the energy density, the initial energy density $\bra{\Psi}T_{00}\ket{\Psi}$  should be $C'$ since the energy conservation. For general operator $O$ we could only fix $\bra{\Psi}O(w_1,\bar w_1)\ket{\Psi}=g(w_1-\bar w_1)$, where $g(w_1-\bar w_1)$ is an arbitrary function of $w_1-\bar w_1$ by considering the translation invariance of $\ket{\Psi}$. \\
In general the 2-point function $\bra{\Psi(\infty)}O(w_1,\bar w_1)O(w_2,\bar w_2)\ket{\Psi(\infty)}=G(w_1,\bar w_1,w_2,\bar w_2)$. By using the KMS condition we have
\begin{eqnarray}
&&\bra{\Psi(\infty)}O(w_1,\bar w_1)\tau_{i\beta}[O(w_2,\bar w_2)]\ket{\Psi(\infty)}\nonumber \\
&&=\bra{\Psi(\infty)}\tau_{-i\beta}\Big[O(w_1,\bar w_1)\tau_{i\beta}[O(w_2,\bar w_2)]\Big]\ket{\Psi(\infty)}\nonumber \\
&&=\bra{\Psi(\infty)}\tau_{-i\beta}[O(w_1,\bar w_1)]O(w_2,\bar w_2)\ket{\Psi(\infty)},
\end{eqnarray}
which suggests that $G(w_1,\bar w_1,w_2,\bar w_2) =G_1(w_1-w_2,\bar w_1-\bar w_2,w_1+\bar w_2,\bar w_1+ w_2)$. The space-translation invariance of state $\ket{\Psi(\infty)}$ implies $G_1(w_1-w_2,\bar w_1-\bar w_2,w_1+\bar w_2,\bar w_1+ w_2)=G_1(w_1-w_2,\bar w_1-\bar w_2,w_1+\bar w_2+2a,\bar w_1+ w_2+2a)$, where $a$ is arbitrary real number.Specially taking $2 a=-(w_1+\bar w_2)$, we would obtain $G(w_1,\bar w_1,w_2,\bar w_2)$ should only be function of $(w_1-w_2)$ and $(\bar w_1-\bar w_2)$, i.e., like the form $G(w_1-w_2,\bar w_1-\bar w_2)$. The KMS condition would constrain function $G(w_1-w_2,\bar w_1-\bar w_2)$ satisfies
\be\label{KMSbeta}
G(w_1-w_2+i\beta,\bar w_1-\bar w_2-i\beta)=G(w_2-w_1,\bar w_2-\bar w_1).
\ee
 For 2D CFT one could obtain the two-point thermal correlation function by a conformal map from complex plane to cylinder with $\tau\sim \tau+\beta$. The thermal two-point correlation function is
 \begin{eqnarray}
 T(w_1-w_2,\bar w_1-\bar w_2)=\Big(\frac{\pi}{\beta }\Big)^{4h} \Big[\sinh(\frac{\pi(w_1-w_2)}{\beta})\sinh(\frac{\pi(\bar w_1-\bar w_2)}{\beta})\Big]^{-2h}.
 \end{eqnarray}
 It is straightforward to check $T(w_1-w_2,\bar w_1-\bar w_2)$ do satisfy the constraint (\ref{KMSbeta})\footnote{If there is no phase transition or spontaneous symmetry breaking,  the KMS state would be unique and equal to the Gibbs state\cite{Araki}. Here we assume this is our case.}.

 If $O(w_1,\bar w_1)$ and $O(w_2,\bar w_2)$ are large (spatially) separated, i.e., $|x_1-x_2|\to \infty$, the cluster property  is expected,\be
\bra{\Psi(\infty)}O(w_1,\bar w_1)O(w_2,\bar w_2)\ket{\Psi(\infty)}\simeq \bra{\Psi(\infty)}O(w_1,\bar w_1)\ket{\Psi(\infty)}\bra{\Psi(\infty)}O(w_2,\bar w_2)\ket{\Psi(\infty)}.
\ee
Thus the connected 2-point function
\begin{eqnarray}\label{timeinfityconnect}
&&C_{t}(x_1,x_2)\equiv \bra{\Psi(t)}O(w_1,\bar w_1)O(w_2,\bar w_2)\ket{\Psi(t)}\nonumber \\
&&- \bra{\Psi(t)}O(w_1,\bar w_1)\ket{\Psi(t)}\bra{\Psi(t)}O(w_2,\bar w_2)\ket{\Psi(t)}\nonumber \\
&&\overset{t\to \infty}{\propto} T(w_1-w_2,\bar w_1-\bar w_2).
\end{eqnarray}
 The connected 2-point function $C_{t}(x_1,x_2)$ at $t$ is expected to be
\be\label{generalform}
C_{t}(x_1,x_2)=M(t)T(w_1-w_2,\bar w_1-\bar w_2)+N(t),
\ee
where $M(t\to\infty)=\text{Constant}\ne 0$ and $N(t\to\infty)=0$, since $T(w_1-w_2,\bar w_1-\bar w_2)$ is invariant under time evolution.
$M(t)$ and $N(t)$ can be seen as function of $w_{1(2)}-\bar w_{1(2)}$ and $w_{1(2)}-\bar w_{2(1)}$. \\
Firstly, let's discuss the form of $N(t)$. Because of the requirement $N(t\to\infty)=0$, we expect $N(t)\sim \frac{1}{t^\alpha}(\alpha>0)$ or $N(t)\sim e^{-\lambda t}(\lambda>0)$ in the large $t$ limit. Without loss of generality we take $w_{1(2)}-\bar w_{1(2)}$  and denote the distance $d(x_1,x_2)=L$ at $t=0$. If $N(t)\sim \frac{1}{t^\alpha}(\alpha>0)$, $N(t)$ can be seen as a function of $w_{1(2)}-\bar w_{2(1)}$,i.e., $N(t)\sim \frac{1}{(w_{1(2)}-\bar w_{2(1)})^\alpha}$. At $t=0$ we would have $C_{t=0}(x_1,x_2)\sim \frac{1}{L^\alpha}$. In this case the correlator in the initial state could be polynomially decaying. If $N(t)\sim e^{-\lambda t}$, at $t=0$, $N(t)\sim e^{-\lambda L}$. In this case $N(t)$ would rapidly decay to zero at the time scale $t\sim 1/\lambda$. For 4D scalar field the correlator the connected 2-point function (\ref{twoscalarcorrelationtime}) can be written as a sum of time-independent part (second line of (\ref{twoscalarcorrelationtime})) and time-dependent part (third line of (\ref{twoscalarcorrelationtime})).\\
In the following let's discuss the possible forms of $M(t)$. Let's first show it would be not possible that $M(t=0)\sim \frac{1}{L^{\alpha'}}(\alpha'>0)$. Since in this case we would have $M(t)\sim \frac{1}{(w_1-\bar w_2-2t)^{\alpha'}}$ at time $t$, which will approach to zero in the limit $t\to \infty$. This is inconsistent with the condition  (\ref{timeinfityconnect}),  $C_{t\to \infty}=T(w_1-w_2,\bar w_1-\bar w_2)$.
If $M(t=0)$ is a finite  constant, and independent with $d(x_1,x_2)$, for $|x_1-x_2|\gg \beta$, we have
\be
C_{t=0}(x_1,x_2)\sim M(t=0) e^{-2 \pi h L/\beta}+...\;.
\ee
It is also possible $M(t=0)\sim e^{-2 \pi L/\beta'}=e^{-2 \pi (w_1-\bar w_2)/\beta'}$, where $\beta'$ is some positive constant. Thought this term blows up in the limit $t\to \infty$, a possible term, such as $e^{2\pi(w_1-\bar w_1)/ \beta'}$(it is $1$ at $t=0$), could cancel the divergence.  As a result the connected 2-point function
\be
C_{t=0}(x_1,x_2)\sim  e^{-2 \pi L/\beta'}e^{-2 \pi h L/\beta},
\ee
for $L\gg \beta$.

In summary  the connected 2-point correlation function in the initial pure state $\ket{\Psi}$ could be exponentially decaying or polynomially decaying . If we write the correlator in the form (\ref{generalform}), $M(t=0)$ would always exponentially decay or a constant for large spatial separation, while $N(t=0)$ could be exponentially decaying or polynomially decaying depending on the details of the states. The boundary state is an example that the initial state has short distance correlation. It would be interesting to find some states which have long distance correlation, but finally would be like a thermal state. \\
Let's briefly discuss the higher dimension theory.
To see the difference between 2D and 4D theories, let's recall the  correlation function of 4D free scalar field (\ref{thermal4D}), which satisfies the KMS condition,
\begin{eqnarray}
T_4(x,y):=\frac{1}{(2\pi)^3}\int \frac{d^3k}{2E_k}\frac{1}{e^{4\tau_0 E_k}-1}\Big(e^{4\tau_0E_k}e^{iE_k(t_y-t_x)}+e^{-iE_k(t_y-t_x)}\Big)e^{i\bm{k}\cdot (\bm{x-y})}
\end{eqnarray}
For $t_x=t_y=0$ we have
\begin{eqnarray}
&&T_4(x,y)=\frac{1}{(2\pi)^3}\int \frac{d^3k}{2E_k} \coth (2\tau_0 E_k) e^{i\bm{k}\cdot (\bm{x-y})}\nonumber \\
&&\quad \quad \quad \quad \sim \frac{1}{r} \coth \big(\frac{\pi r }{4 \tau_0}\big)\;\;{\sim}\;\; \frac{1}{r},
\end{eqnarray}
for $r\gg \tau_0$. This is different from the 2D thermal correlation function, which is exponentially decaying for large spatial separation. So our discussion based on 2D thermal correlation function would break down in higher dimension.

\subsection{Local thermalization}

When discussing the entanglement property of regularized boundary state, we assume the parameter $\tau_0$ is very small comparing with the distance $d(x_1,x_2)$ between operators. In this section we consider the opposite limit, i.e., the distance  between operators $d(x_1,x_2)\ll \tau_0$. This limit
is closely related to the (local) eigenstate thermalization hypothesis (ETH). Specially for 2D CFT with large central charge in paper  \cite{Fitzpatrick:2014vua}\cite{Fitzpatrick:2015zha} the authors show the 2-point function of light operators is consistent with the ones in a highly excited pure state at the first order of central charge. Recently there are many processes on local ETH by using other physical quantities\cite{Lashkari:2016vgj}-\cite{Basu:2017kzo}, such as entanglement entropy, relative entropy, etc. \\

Let's start with the 2-point function (\ref{2D2point}). We would like to consider the limit $d(x_1,x_2)\ll \tau_0$, which will lead to the cross ration $x\to 1$. The 2-point function becomes the thermal 2-point function (\ref{Thermal2point}), i.e., in the limit $\tau_0\gg d(x_1,x_2)$
\be\label{localthermai}
~_a^\t0\!\bra{B}O(w_1,\bar w_1)O(w_2,\bar w_2)\ket{B}^{\tau_0}_{a}\to\Big(\frac{\pi}{4\tau_0}\Big)^{4h}\Big[\sinh(\frac{\pi(w_1-w_2)}{4\tau_0})\sinh(\frac{\pi(\bar w_1-\bar w_2)}{4\tau_0})\Big]^{-2h}.
\ee
By similar argument one could obtain the 3-point function would approach the thermal 3-point function (\ref{thermal3point}) in this limit.
The result is simple but the meaning of $\tau_0$ is still not clear. $\ket{B}^{\tau_0}_a$ is not an eigenstate of $H$, but a superposition of different energy eigenstates. (\ref{cardystate}) shows the Cardy states are linear combinations of Ishibashi states, conversely we have
\be\label{Ishibashiinverse}
\ket{i}\rangle=\sum_{a}S^i_a \sqrt{S^i_0} \ket{B}_a.
\ee
We still regularize the Ishibashi state as $\ket{i}\rangle^{\tau_0}\equiv e^{-\tau_0H}\ket{i}\rangle$. To obtain the 2-point function in the Ishibashi state we need
\be
~_a^\t0\!\bra{B}O(w_1,\bar w_1)O(w_2,\bar w_2)\ket{B}^{\tau_0}_{b}.
\ee
This correlator can be seen as correlation function on a strip of width $2\tau_0$, with boundary conditions on $\tau_E=\tau_0,-\tau_0$ respectively corresponding to boundary state $\ket{B}^{\tau_0}_{b}$ and $\ket{B}^{\tau_0}_{a}$. With the conformal mapping (\ref{conformalmap}) the strip is mapped to UHP, but imposed different boundary condition on the positive and negative real axis, which corresponds to the insertion of a boundary operator $\phi_{ab}(0)$\cite{Cardy1}. In paper \cite{Cardy:2017ufe} the author argues the norm $Z_{ab}\equiv ~_a^{\tau_0}\!\bra{B}e^{-2\tau_0 H}\ket{B}_{b}^{\tau_0}$ ($a\ne b$) should be much smaller than the case $a=b$ if the length of the strip $R\gg \tau_0$, which permits us to write $Z_{ab}=\delta_{ab} Z_{aa}$. Similarly, the correlators
\be\label{twopointdifferntboundary}
~_a^\t0\!\bra{B}O(w_1,\bar w_1)O(w_2,\bar w_2)\ket{B}^{\tau_0}_{b}\simeq \delta_{ab}~_a^\t0\!\bra{B}O(w_1,\bar w_1)O(w_2,\bar w_2)\ket{B}^{\tau_0}_{a}.
\ee
One could understand this result as follows. The image of $x_1,x_2$ under the conformal mapping (\ref{conformalmap}) is far away from the origin in UHP. According to the clustering property
\begin{eqnarray}
&&\langle O(w_1,\bar w_1)O(w_2,\bar w_2) \phi_{ab}(0)\rangle_{\text{UHP}}\nonumber \\
&&\simeq \langle O(w_1,\bar w_1)O(w_2,\bar w_2)\rangle_{\text{UHP}} \langle\phi_{ab}(0)\rangle_{\text{UHP}}\nonumber \\
&&= \langle O(w_1,\bar w_1)O(w_2,\bar w_2)\rangle_{\text{UHP}} Z_{aa}\delta_{ab}.
\end{eqnarray}
By using (\ref{Ishibashiinverse}) and (\ref{twopointdifferntboundary}) we obtain
\begin{eqnarray}
~^{\tau_0}\!\langle \bra{i} O(w_1,\bar w_1)O(w_2,\bar w_2) \ket{i}\rangle^{\tau_0}= \sum_{a}(S_{a}^i)^*S_{a}^i\sqrt{S^i_0(S^i_0)^*} \nonumber ~_a^\t0\!\bra{B}O(w_1,\bar w_1)O(w_2,\bar w_2)\ket{B}^{\tau_0}_{a}.
\end{eqnarray}
In the limit $\tau_0\gg d(x_1,x_2)$ we could obtain the 2-point function in Ishibashi state, which is also a thermal 2-point correlation function (\ref{Thermal2point}), because $~_a^\t0\!\bra{B}O(w_1,\bar w_1)O(w_2,\bar w_2)\ket{B}^{\tau_0}_{a}\to T(w_1-w_2,\bar w_1-\bar w_2)$ is independent on the boundary condition $a$. Although this result is derived by using (\ref{cardystate}), which is usually true for minimal models, we expect it is also true for more general conformal field theory, e.g., the large $c$ CFTs. \\

Assume the strip is made periodic in the $x$-direction with radius $R\gg \tau_0$, and the Hamiltonian $H$ for the cylinder is related to the Virasoro generators $L_0$ and $\bar L_{0}$ on z-plane by the conformal mapping (\ref{cylindertoz}),
\be
H=\frac{1}{R}(L_0+\bar L_0-\frac{c}{12}).
\ee
An Ishibashi state $\ket{i}\rangle^{\tau_0}$  is a superposition of states in the $i$-th Verma modules (\ref{Ishibashistate}), so the regularized Ishibashi state is
\be
\ket{i}\rangle^{\tau_0}= e^{-\tau_0H}\ket{i}\rangle=\sum_N e^{-\frac{\tau_0}{R}(h+\bar h+2N-c/12)}\ket{i+N}\otimes\ket{\bar i+N},
\ee
with $i=\bar i$. The higher energy contribution $N\gg h$ is suppressed, the regularized Ishibashi state can be  effectively seen as a superposition of
finite number of energy eigenstates. When $\tau_0$ is large, the leading term is the highest weight vector $\ket{i}\otimes \ket{\bar i}$. For state $\ket{i}\rangle^{\tau_0}$ the energy density $\langle T_{tt}\rangle\sim \frac{c}{\tau_0^2}$, thus the energy $E\sim \frac{c R}{\tau_0^2}$. If only consider the leading term in the state $\ket{i}\rangle^{\tau_0}$,  we may estimate $\tau_0\sim ch/E_h$, where $E_h$ is the energy of $\ket{i}\otimes \ket{\bar i}$. The above argument implies the thermal property (\ref{localthermai}) also appears in the pure state $\ket{i}\otimes \ket{\bar i}$ in the limit $d(x_1,x_2)\ll ch/E_h$.

\section{Conclusion}
In this paper we discuss some properties on the regularized boundary state $\ket{B}^{\tau_0}_{a}$ (\ref{regularizedstate}). In 2D CFT the correlation function in this state is always exponential decay, which hints there exists an energy gap. This permits us to derive a cluster property for bounded operators. The correlation strength is directly related to quantum entanglement. By using Bell inequality as a witness we obtain that the quantum entanglement is exponential decay under large spatial separation (\ref{estimationBell}), which is quite different from the vacuum of CFT.  This upper bound is similar as the one in vacuum state of a theory with mass gap, implying some relation between the regularized boundary state $\ket{B}^{\tau_0}_{a}$ and ground state of a CFT deformed by relevant bulk operators\cite{Cardy:2017ufe}.\\
When taking $\ket{B}^{\tau_0}_{a}$ as an initial state in a CFT, it will evolute under the Hamiltonian of this theory. After long time the state would exhibit thermal property. We show this both in a free field theory in higher dimension and 2D CFT. We use the KMS condition to characterize the thermal property of the finial state. The 2-point correlation functions of local operators do satisfy the KMS condition with $\beta=4\tau_0$. We also discuss the higher point correlation functions by some special examples. \\
Actually the initial state $\ket{B}^{\tau_0}_{a}$ also partially  satisfies the KMS condition (\ref{KMS}), but the function $F(z)=\omega(A\tau_{z}(B))$ is non-analytical. The role of time evolution is just to move the poles to infinity and make $F(z)$ be a analytic function on region $0<\beta<4\tau_0$. This is just a mathematical view on the role of time evolution, the physical meaning behind which is still not clear.  We generally analyse the pure state quantum quench process and find some constraints on the initial state if asking the final state locally approach to a KMS state. We discuss the possible form of correlation function in the initial state.  \\
As a byproduct we find in an opposite limit, i.e., $\tau_0\gg d(x_1,x_2)$, 2-point function in $\ket{B}^{\tau_0}_{a}$ will behave like a thermal 2-point function. This is also true for Ishibashi state, which can be seen as a superposition of different energy in a Verman modules $\mathcal{V}_i\bigotimes \mathcal{\bar V}_{\bar i}$. In the large $\tau_0$ limit the leading term would be the highest weight state $\ket{i}\otimes \ket{i}$. The thermal property can be seen as in the pure state $\ket{i}\otimes \ket{i}$ with the limit that $d(x_1,x_2)\ll ch/E_h$.

\acknowledgments

I would like to thank Chong-Sun Chu and Feng-Li Lin for useful discussion and carefully reading the manuscript.  I am thankful to Feng-Li Lin for his encouragement and giving many helps during my visiting at NTNU. I also would like to thank the referee for continuous discussion and encouragement.  This work is supported in part by the National Center of Theoretical Science (NCTS).

\appendix
\section{Four point function of creation and annihilation operators}\label{4point}
The following is the detail on the calculation of four-point correlation function \\ $~_{\pm}^\t0\!\bra{B}a^\dagger_{\bm{p}_2}a^\dagger_{\bm{p}_1}a_{\bm{k}_1}a_{\bm{k}_2}\ket{B}^{\tau_0}_{\pm}$. By the definition and communication relation, we have
\be
a_{\bm{k}}\ket{B}^{\tau_0}_{\pm}=\pm e^{-2\tau_0 E_k} a^{\dagger}_{\bm{-k}}\ket{B}^{\tau_0}_{\pm}, \quad \text{and} \quad ~_{\pm}^\t0\!\bra{B}a^\dagger_{\bm{p}}=\pm e^{-2\tau_0 E_p}~_{\pm}^\t0\!\bra{B}a_{\bm{-p}}.
\ee
Using these relation we obtain
\begin{eqnarray}
&&~_{\pm}^\t0\!\bra{B}a^\dagger_{\bm{p}_2}a^\dagger_{\bm{p}_1}a_{\bm{k}_1}a_{\bm{k}_2}\ket{B}^{\tau_0}_{\pm}\nonumber \\
&&=e^{-2\tau_0 (E_{p_2}+E_{k_2})}\Big( \delta(\bm{p_1}+\bm{p_2})\delta(\bm{k_1}+\bm{k_2})\frac{e^{4\tau_0E_{p_2}}e^{4\tau_0E_{k_2}}-1}{(e^{4\tau_0E_{p_2}}-1)(e^{4\tau_0E_{k_2}}-1)}\nonumber \\
&&+\delta(\bm{p_1}-\bm{k_1})\delta(\bm{p_2}-\bm{k_2})\frac{1}{e^{4\tau_0E_{p_1}}-1}\Big)
+e^{-2\tau_0 (E_{p_2}+E_{k_2})}~_{\pm}^\t0\!\bra{B}a^\dagger_{-\bm{k}_2}a^\dagger_{\bm{p}_1}a_{\bm{k}_1}a_{-\bm{p}_2}\ket{B}^{\tau_0}_{\pm},\nonumber \\
\end{eqnarray}
one could calculate the last term by taking $\bm{p}_2\to -\bm{k}_2$ and $\bm{k}_2\to -\bm{p}_2$ in the original expression. Finally the result is
\begin{eqnarray}
&&~_{\pm}^\t0\!\bra{B}a^\dagger_{\bm{p}_2}a^\dagger_{\bm{p}_1}a_{\bm{k}_1}a_{\bm{k}_2}\ket{B}^{\tau_0}_{\pm}\nonumber \\
&&=e^{-2\tau_0 (E_{p_2}+E_{k_2})}\Big( \delta(\bm{p_1}+\bm{p_2})\delta(\bm{k_1}+\bm{k_2})\frac{e^{4\tau_0E_{p_2}}e^{4\tau_0E_{k_2}}-1}{(e^{4\tau_0E_{p_2}}-1)(e^{4\tau_0E_{k_2}}-1)}\Big)\\ \nonumber
&&+\delta(\bm{p}_1-\bm{k}_1)\delta(\bm{p}_2-\bm{k}_2)\frac{e^{-4\tau_0 E_{p_2}}+e^{-8\tau_0 E_{p_2}}}{e^{4\tau_0 E_{p_1}}-1}+ \delta(\bm{p}_1-\bm{k}_2)\delta(\bm{p}_2-\bm{k}_1)\frac{1-e^{-4\tau_0 (E_{p_2}+E_{k_2})}}{(e^{4\tau_0E_{p_2}}-1)(e^{4\tau_0E_{k_2}}-1)} \\ \nonumber
&&+e^{-4\tau_0(E_{k_2}+E_{p_2})}~_{\pm}^\t0\!\bra{B}a^\dagger_{\bm{p}_2}a^\dagger_{\bm{p}_1}a_{\bm{k}_1}a_{\bm{k}_2}\ket{B}^{\tau_0}_{\pm}.
\end{eqnarray}
We could solve $~_{\pm}^\t0\!\bra{B}a^\dagger_{\bm{p}_2}a^\dagger_{\bm{p}_1}a_{\bm{k}_1}a_{\bm{k}_2}\ket{B}^{\tau_0}_{\pm}$,
\begin{eqnarray}
&&~_{\pm}^\t0\!\bra{B}a^\dagger_{\bm{p}_2}a^\dagger_{\bm{p}_1}a_{\bm{k}_1}a_{\bm{k}_2}\ket{B}^{\tau_0}_{\pm}\nonumber \\
&&=\delta(\bm{p_1}+\bm{p_2})\delta(\bm{k_1}+\bm{k_2}) \frac{e^{2\tau_0(E_{k_2}+E_{p_2})}}{(e^{4\tau_0E_{p_2}}-1)(e^{4\tau_0E_{k_2}}-1)}\nonumber \\
&&+\Big(\delta(\bm{p}_1-\bm{k}_1)\delta(\bm{p}_2-\bm{k}_2)+\delta(\bm{p}_1-\bm{k}_2)\delta(\bm{p}_2-\bm{k}_1)\Big) \frac{1}{(e^{4\tau_0E_{p_1}}-1)(e^{4\tau_0E_{p_2}}-1)}.
\end{eqnarray}
\section{The proof of the theorem in section \ref{KMSconditon} } \label{theorem}
The theorem:\\

\emph{If $\omega$ is a $(\tau,\beta)$-KMS state, with $\beta \ne 0$, it follows that $\omega$ is time-invariant, i.e.,
\be
\omega(\tau_t (A))=\omega(A),
\ee
for all operators $A$ and $t\in \mathbb{R}$.
}
~\\
Proof.  For operators $A$ define the analytic function $F$ by
\be
F(z)=\omega(\tau_z(A)).
\ee
Define
\be M=\text{sup}\{\tau_{i\gamma}(A),\gamma\in [0,\beta]\}.\ee
We have
\be |F(z)|\le \Vert \tau_z(A)\Vert =\Vert \tau_{\text{Re}\  z}(\tau_{i\  \text{Im} z}(A))\Vert=\Vert \tau_{i\text{Im} z}(A)\Vert\le M. \ee
For $I$ and $A$ it follows directly from the KMS relation that
\be F(z+i\beta)=\omega (I \tau_{i\beta}(\tau_z (A)) )=\omega (\tau_z (A))=F(z),\ee
which means $F(z)$ is a periodic function with period $i\beta$. For all the $z\in \mathcal{C}$ this implies
\be |F(z)|\le M.\ee
Hence $F(z)$ is a constant by Liouville's theorem.

\paragraph{}


\end{document}